\documentclass[aps,preprint,showpacs]{revtex4}
\usepackage{epsfig}
\usepackage{amssymb}
\usepackage{multirow}
\usepackage{dcolumn}
\usepackage{bm}
\begin{document}

\title{Quasinormal modes of a Generic-class of magnetically charged regular black hole: scalar and electromagnetic perturbations}

\author{L. A. L\'opez}
\email{lalopez@uaeh.edu.mx}
\author{Valeria Ram\'irez} 
\email{ra323273@uaeh.edu.mx}

\affiliation{ \'Area Acad\'emica de Matem\'aticas y F\'isica, UAEH, 
Carretera Pachuca-Tulancingo Km. 4.5, C P. 42184, Mineral de la Reforma, Hidalgo, M\'exico.}

\begin{abstract}
In this contribution, we study the quasinormal modes of a Generic--class of a regular black hole with a magnetic charge in nonlinear electrodynamics, considering scalar and electromagnetic perturbations. The Generic--class contains the Bardeen--class, Hayward--class, and  New--class solutions. As the Generic--class can represent a black hole with two horizons or one horizon. First, we obtain the critical values of the magnetic charge and mass. Then, using the third--order WKB approximation, we can determine the dependence of the quasinormal modes with the parameters of the Generic--class. Finally, the transmission and reflection coefficients of the scattered wave in the third--order WKB approximation are calculated.
\end{abstract}

\maketitle

\section{Introduction}

In general, the solutions to the Einstein equations with spherical symmetry representing Black Holes (BHs) have a singularity at the origin. According to the Penrose conjecture, these singularities must be dressed by event horizons, the singularities are nonphysical objects so they do not exist in nature. The violation of the Penrose conjecture is produced when instabilities exist, leading to the destruction of the event horizon and producing naked singularities. 

The construction of regular solutions (free of curvature divergences) has been proposed to avoid the singularity problem, the theory of general relativity coupled to nonlinear electrodynamics has been a good candidate. Born and Infeld \cite{Born:1934gh} were the first ones that considered the nonlinear electrodynamics as an attempt to avoid, at a classical level, the singularity of the electric field of a point charge.This is why the regular magnetic black hole proposed by Bardeen \cite{Ayon-Beato:2000mjt} exists. 
Another regular solution that contains critical scale, mass, and charge parameters was proposed by Hayward \cite{Hayward:2005gi}, this is accomplished by applying the limiting curvature conjecture \cite{Polchinski:1989ae}.

A Generic-class that contains the solutions of Bardeen, Hayward, and another one is addressed in \cite{PhysRevD.94.124027} where a regular solution of a charged black hole in general relativity coupled to nonlinear electrodynamics (NLED) is constructed by the authors.

A black hole constantly interacts with matter and fields that surround it, as a result of these interactions, the BH is in a perturbed state, the perturbations of the black hole are independent of whether the BH is regular or irregular. When a BH is perturbed, the resulting behavior can be described as a stage of damped oscillations with complex frequencies, the modes of such oscillations are called quasinormal modes (QNM). The real part of the frequencies describes the frequency of oscillation, and the imaginary part describes the damping of the oscillation. 

An application of the study of QNM is the analysis of the stability of BHs, they are also crucial for characterizing the gravitational wave signals detected by LIGO  and VIRGO \cite{LIGOScientific:2016sjg}. It is important to mention that several numerical methods to calculate the QNM exist; for example, the continued fraction method \cite{Percival:2020skc}, finite difference method \cite{Ma:2020qkd}, WKB  \cite{Schutz:1985km}, among other approximation methods, being the WKB method the most used.

Nowadays, different authors have investigated QNM in NLED. For instance, the QNM of Bardeen BH \cite{PhysRevD.86.064039},  QNM of charged BHs in Einstein--Born--Infeld gravity \cite{Fernando:2004pc}, Bronnikov BH \cite{Li:2014fka}, among others. Scalar, electromagnetic and gravitational perturbations are considered in \cite{Toshmatov:2018ell} and \cite{Toshmatov:2018tyo}. In addition, the QNM of Hayward, Bardeen, and Ayón--Beato--García are compared in \cite{Toshmatov:2015wga}. QNM may also be studied using the Eikonal regime and effective geometry as it can be seen in \cite{Breton:2016mqh} and in \cite{PhysRevD.104.024064} where the Eikonal approximation is applied to study the Einstein--Euler--Heisenberg BH.

Moreover, different investigations have emerged about QNM for other types of scenarios. This can be illustrated in a couple of references, in \cite{Dey:2020lhq} Braneworld black holes are studied, as well as in \cite{Bronnikov:2012ch}\cite{Franzin:2022iai} where black holes and wormholes are considered. 

The structure of this paper is organized as follows: The subsequent section presents an analysis of the Generic--class of magnetically charged regular BH, the conditions for it to have one or two event horizons are given. Section III derives the effective potentials arising in the scalar and electromagnetic perturbations. Then, the behavior of the potentials for the Bardeen--class, Hayward--class, and a New--class solution is shown. In  Section IV presents the QNM of the scalar and electromagnetic perturbations. The frequencies of the QNM are compared considering different values of the constant that characterizes the strength of nonlinearity of the electromagnetic field. Finally, the Section V includes the reflection and transmission coefficients considering the different perturbations. Conclusions are given in Section VI.

\section{Regular Black Holes in NLED}

The general action that describes the gravity coupled to nonlinear electrodynamics \cite{PhysRevD.61.045001} is given by;

\begin{equation}\label{action}
S=\frac{1}{16\pi}\int d^{4}x\sqrt{-g}[R-L(F,G)],
\end{equation}

where $R$ is the scalar of curvature. $L$ is an arbitrary function of  electromagnetic invariants $F=F_{\beta \lambda}F^{\beta \lambda}$ and $G=F_{\beta \lambda}\mathcal{F}^{\beta \lambda}$, where $F_{\beta \lambda}=\partial_{\beta}A_{\lambda}-\partial_{\lambda}A_{\beta}$ is the electromagnetic field tensor and $\mathcal{F}^{\beta\lambda}$ is the dual field electromagnetic tensor. This paper considers that the Lagrangians only depend on $F$. The energy momentum tensor is;

\begin{equation}
T_{\beta \lambda}= 2(L_{F}F^{\gamma}_{\beta} F_{\lambda \gamma}-\frac{1}{4}g_{\beta \lambda}L),
\end{equation}

the subindex $F$ in $L$ represents the derivative with respect to $F$.

In spherical coordinates $x^{\mu}=(t,r,\theta,\phi)$, we can consider a static, spherically symmetric line element;

\begin{equation}\label{sss}
ds^{2}=-f(r)dt^{2}+\frac{1}{f(r)}dr^{2}+r^{2}d\Omega^{2},
\end{equation} 

with $f(r)=1-m(r)/r$, where $m(r)$ represents a finite-mass distribution function. To exclude the singularity at the origin is necessary to consider some restrictions: the function $m(r)$ is at least three times differentiable, it approaches zero sufficiently fast in the limit $r \to 0$, and the third-order derivative has some low--lying curvature polynomials (for more details see \cite{PhysRevD.94.124027}).

The line element (\ref{sss}) satisfies the symmetry $T^{t}_{t}=T^{r}_{r}$ and the potential can be written as;

\begin{equation}
A_{\beta}=\tilde{ \varphi} (r)\delta^{t}_{\beta}-q\cos \theta \delta^{\phi}_{\beta}
\end{equation}

Where $\tilde{\varphi} (r)$ is the electric potential and the magnetic charge is represented by $q$.

The metric funtion of the Generic--class of magnetically charged regular BH (see \cite{PhysRevD.94.124027}) is given by;

\begin{equation}\label{fss}
f(r)=1-\frac{2Mr^{\mu-1}}{(r^{\nu}+q^{\nu})^{\mu / \nu}}
\end{equation}

The dimensionless constant that characterizes the electromagnetic field's nonlinearity strength is given by $\mu \geq 3$, $M$ is the gravitational mass, and $\nu >0$ is a dimensionless constant. The form the Lagrangian density to obtain $f(r)$ in (\ref{fss}) is given by;

\begin{equation}
L=\frac{4\mu}{\alpha}\frac{(\alpha F)^{\frac{\nu +3}{4}}}{[1+(\alpha F)^{\frac{\nu}{4}}]^{1+\frac{\mu}{\nu}}}.
\end{equation}

$\alpha > 0$ is a constant that has the dimension of length squared and it is related to the gravitational mass ($M=\alpha^{-1}q^{3}$). Moreover, the construction of charged BHs has been studied in \cite{PhysRevD.96.128501} \cite{PhysRevD.98.028501}.

From (\ref{fss}) several classes of regular BHs can be obtained, such as Bardeen--class (Barc) solutions ($\nu = 2$), Hayward--class (Hawc) solutions ($\nu= \mu$) and a New--class (Newc) ($\nu = 1$).  The solution (\ref{fss}) can represent a BH with two horizons, or an extremal BH with one horizon. The number of horizons depends on the possible values of parameters $\mu$, $\nu$, $M$, and $q$.

The event horizons ($r_{h}$) for the space--time defined by line element (\ref{sss}) are obtained by obtaining the roots of the function $f(r)$,  only for this analysis, the radial distance and the charge parameter $q$ are expressed in units of mass as $r \to r/M$,  $q \to q /M $.

As a next step, we determine the range of values of  magnetic charge $q$,  were the line element  (\ref{sss}) can represent a black hole or an extremal black hole, for this analysis, the method described in \citep{Toshmatov:2015npp,PhysRevD.98.024015} id used. First, we consider that $f(r_{h})=0$, then  $q^{\nu}$  can be parametrized as a function of $r_{h}$ as;

\begin{equation}\label{qnu}
q^{\nu}(r_{h})= -r_{h}^{\nu}+\frac{2^{\nu / \mu}}{(r_{h}^{1-\mu})^{\nu / \mu}}
\end{equation}

where $q^{\nu}(r_{h})$ has extremes $\left (\frac{d q^{\nu}(r_{h})}{dr_{h}}=0\right )$ in a certain $r_{h}=r_{cri}$, given by; 

\begin{equation}
r_{crit}= \left (\frac{2^{\nu / \mu}(\mu -1)}{\mu} \right )^{\mu / \nu}
\end{equation}

Then, the critical value of the parameter $q^{\nu}$ is given in terms of $\mu$ as; 

\begin{equation}
q_{crit}^{\nu}= q^{\nu}(r_{crit})=\frac{\mu  \left(\frac{(\mu -1) 2^{\nu /\mu }}{\mu }\right)^{\mu /\nu }}{\mu -1}-\left(\frac{(\mu -1) 2^{\nu /\mu }}{\mu }\right)^{\mu },
\end{equation}

so the magnetically charged regular BH (\ref{fss}) has one ($r_{+}$) or two ($r_{in}$ and $r_{out}$) horizons for values of $q^{\nu}\leq q_{crit}^{\nu}$.  

When the equations $f(r)=0$ and $\frac{d}{dr}f(r)=0$ are satisfied simultaneously, the extreme case (the inner and outer horizons form a only horizon $r_{+}$) is obtained. Then, introducing  $q^{\nu}$ of  (\ref{qnu}) in $\frac{d}{dr}f(r)=0$, we can obtain one real root denoted by $r_{+}=r_{crit}$.

\begin{figure}[ht]
\begin{center}
\includegraphics [width =0.4 \textwidth ]{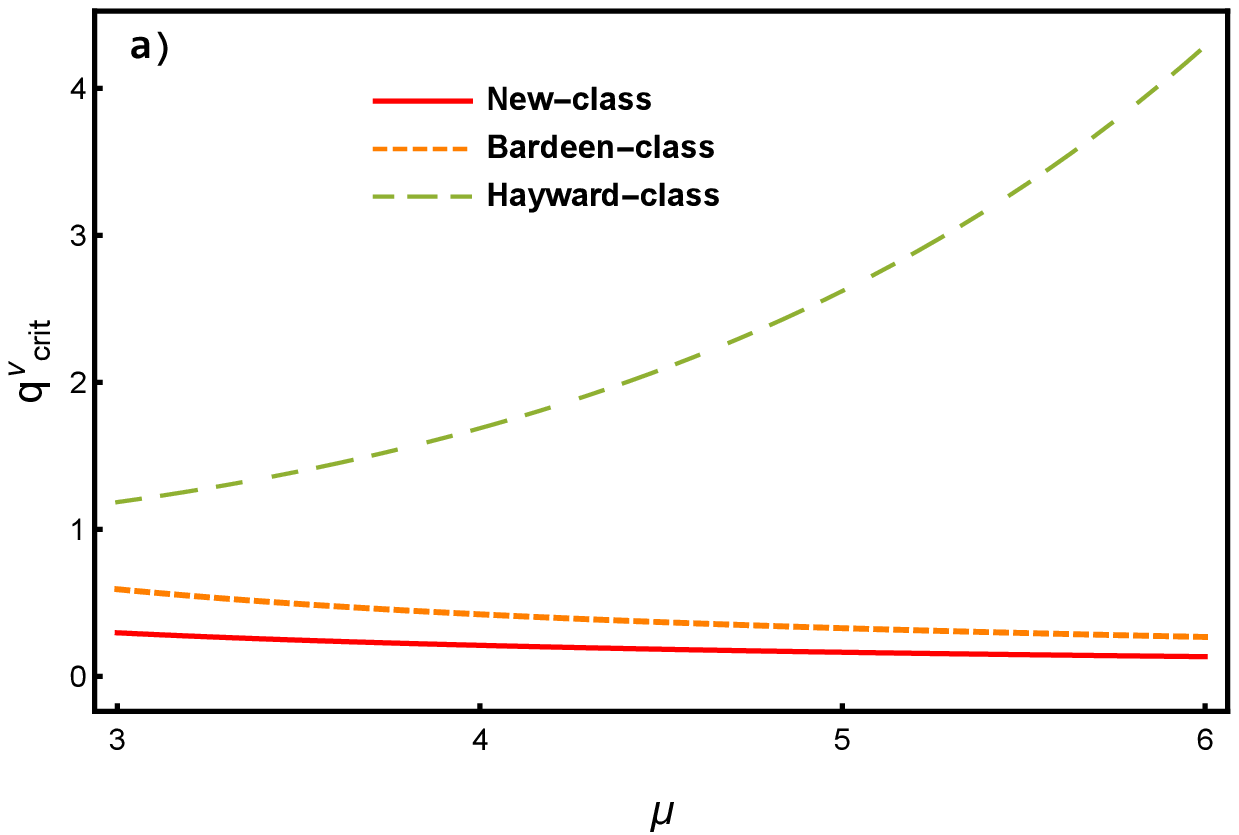}
\includegraphics [width =0.41 \textwidth ]{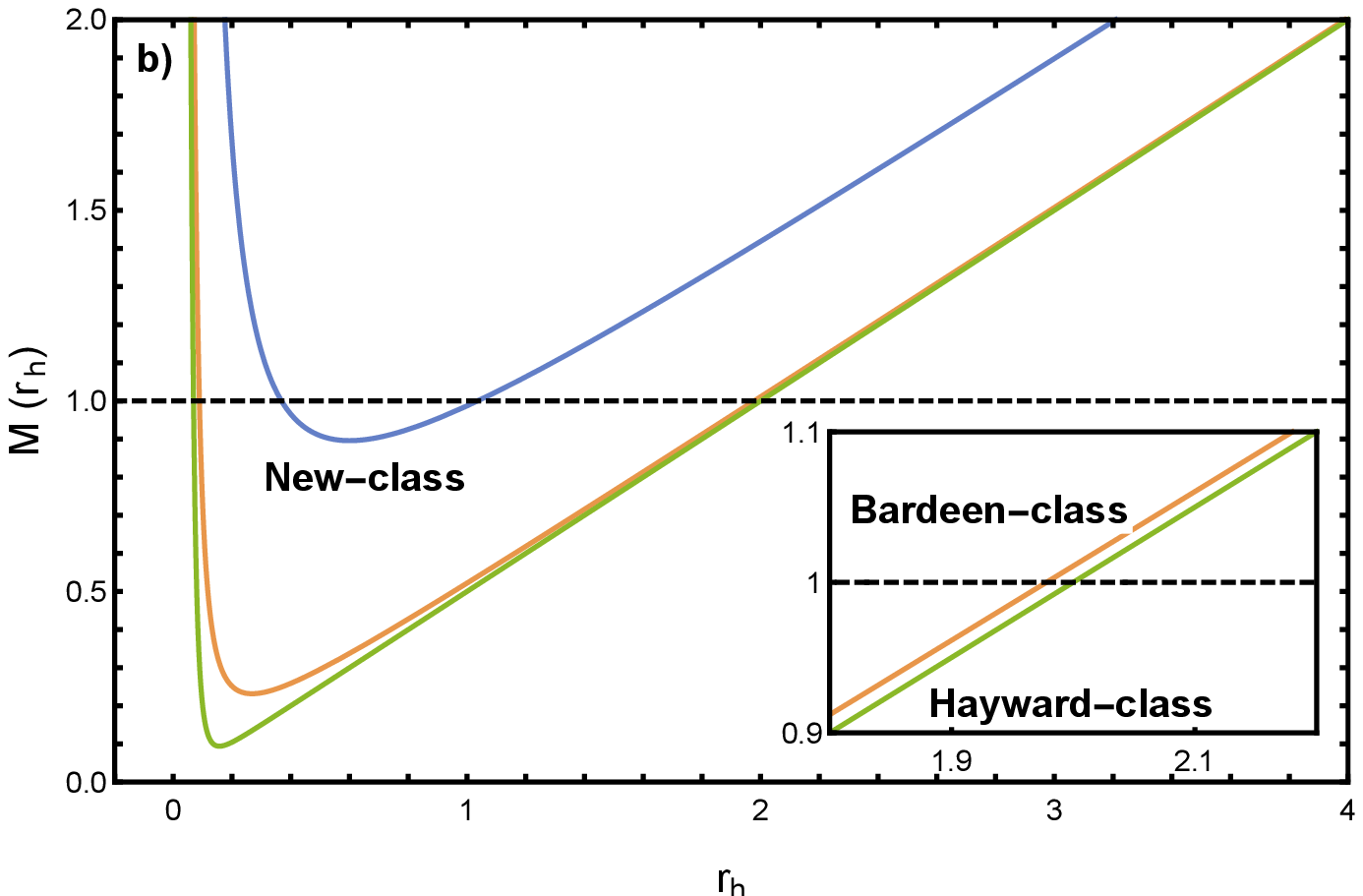}
\end{center}
\caption{The figure a) shows the behavior of $q^{\nu}_{crit}$ in function of $\mu$. Notice that as $\mu$ grows, $q^{\nu}_{crit}$ also grows only in the case of Hawc. The figure b) shows the behavior of $M$ as function of $r_{h}$, with $\mu = 6$ and $q=0.12$}\label{Fig1}
\end{figure}

The Fig. \ref{Fig1} a) shows the behavior of $q^{\nu}_{crit}$ as function of $\mu$. The lines $q^{\nu}_{crit}$ contain the values of ($q^{\nu},\mu$) where the magnetically charged regular black holes have one horizon, while in the regions below the lines, there are two horizons and there are no horizons in the regions above the lines. 

Thus, the ranges of the charge parameter increase as $\mu$ increases in the Hawc solution ($\nu= \mu$) and decrease as $\mu$ increases in the Barc ($\nu = 2$) and in the Newc ($\nu = 1$). It is also possible to mention that the range $[q]_{\nu = \mu}> [q]_{\nu = 2}>[q]_{\nu = 1}$. In \cite{PhysRevD.99.064043} the ranges of $q$ for the case with  $\mu = 3$ were studied.

We complement the analysis of  the horizons of Generic--class, studying the behavior of $M(r_{h},q)$ in (\ref{fss}).

\begin{equation}
M(r_{h},q)=\frac{1}{2}r_{h}^{1 - \mu} (q^{\nu} + r_{h}^{\nu})^{\mu / \nu}
\end{equation}

Note that as $\mu \geq \nu$ the function $M(r_{h},q) \to \infty$  for $r_{h} \to \infty$.  A critical mass $M_{*}$ exists (see equation (\ref{M})) and a radius $r_{*}=q (-1 + \mu)^{1/\nu}$ such that the Generic--class does not have a horizon when $M < M_{*}$, in the case that $M_{*}=M$, the Generic--class has one horizon at  $r=r_{*}$ and has two horizons if $M_{*}<M$.

\begin{equation}\label{M}
M_{*}= \frac{1}{2} q \mu^{\mu / \nu}(\mu-1)^{\frac{1-\mu}{\nu}}
\end{equation}

In Fig.\ref{Fig1} b) the behavior of $M$ as function of $r_{h}$ is shown for $\mu=6$, it can be seen that $M_{*(Hawc)}<M_{* (Barc)}<M_{*(Newc)}$. From the Fig. \ref{Fig1} b) we can conclude that when we fix a value of $M$ (for example $M=1$) is shown, the $r_{out}$ horizon of Hawc is larger compared with $r_{out}$ horizons of Barc and Newc.

\section{Scalar and electromagnetic perturbations}

The Klein--Gordon equation that describe the  perturbation equation for the massless scalar field is given by;

\begin{equation}\label{ec.sfkg}
\frac{1}{\sqrt{-g}}\partial_{\mu}\left(\sqrt{-g}g^{\mu\nu}\partial_{\nu}\right)\Phi=0\,.
\end{equation}
Considering the line element Eq. (\ref{sss}) in the Eq. (\ref{ec.sfkg}), and using the a scalar field $\Phi$ given by;  
\begin{equation}
\Phi=e^{-i\omega t}Y_{l}^{m}(\theta,\phi)\frac{\tilde{\xi}(r)}{r}\,.
\end{equation}

we obtain the radial perturbation equation 
\begin{equation}\label{ecmcnsf}
\frac{d^2\tilde{\xi}(r)}{dr_*^2}+\left[\omega^2-V(r)\right]\tilde{\xi}(r)=0\,,
\end{equation}  

with the tortoise coordinates $dr_*=\frac{dr}{f(r)}$, $\omega$ the frequency, the effective potential $V(r)=f(r)\left[\frac{l(l+1)}{r^2} +\frac{1}{r} \frac{df(r)}{dr}\right]$ and $l$ is the spherical harmonic index of $Y_{l}^{m}(\theta,\phi)$.

When we consider electromagnetic fields, we can generalize the form of the effective potential for scalar and electromagnetic test fields as follows;

\begin{equation}\label{ec.poseg}
V(r)=f(r)\left[\frac{l(l+1)}{r^2} +\left(1-s\right)\left(\frac{1}{r}\frac{df(r)}{dr}
\right) \right] \,,
\end{equation}

Here, $l$ is restricted by $l\geq s$, and $s$ denotes the spin of the scalar ($s=0$) or electromagnetic ($s=1$) perturbations \cite{PhysRevD.71.124033, Konoplya:2011qq, Nollert:1999ji}. The effective potential $V(r)$ (\ref{ec.poseg}) depends on the parameters of the function $f(r)$ and the harmonic index $l$. It may be noted that the value of both effective potentials (see equation \ref{ec.poseg}) increases with an increasing  $l$, then, to be able to make a comparison between the different effective potentials of the different solutions that emerge from (\ref{fss}), we choose a particular case of $l$.  

\begin{figure}[ht]
\begin{center}
\includegraphics [width =0.48 \textwidth ]{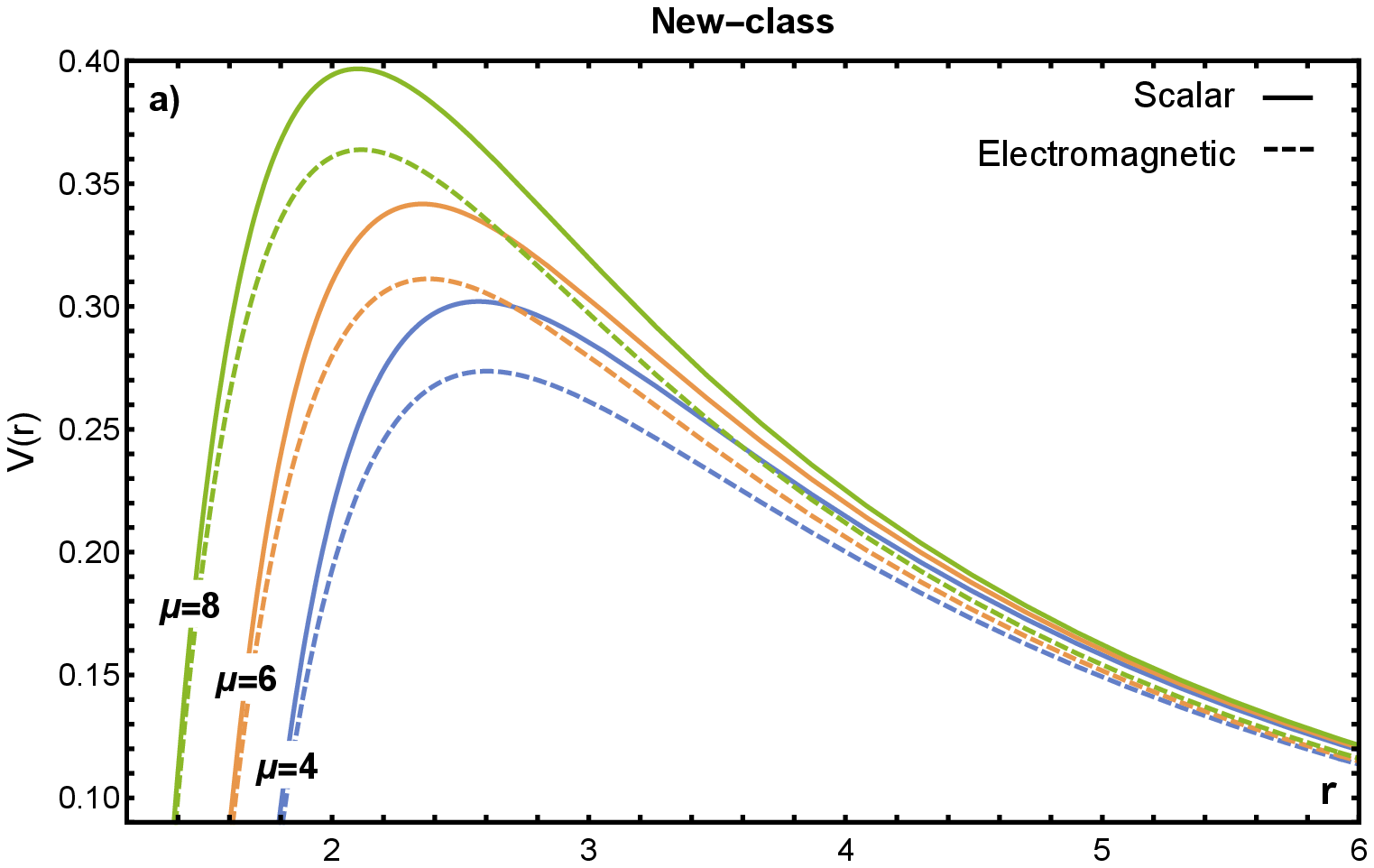}
\includegraphics [width =0.48 \textwidth ]{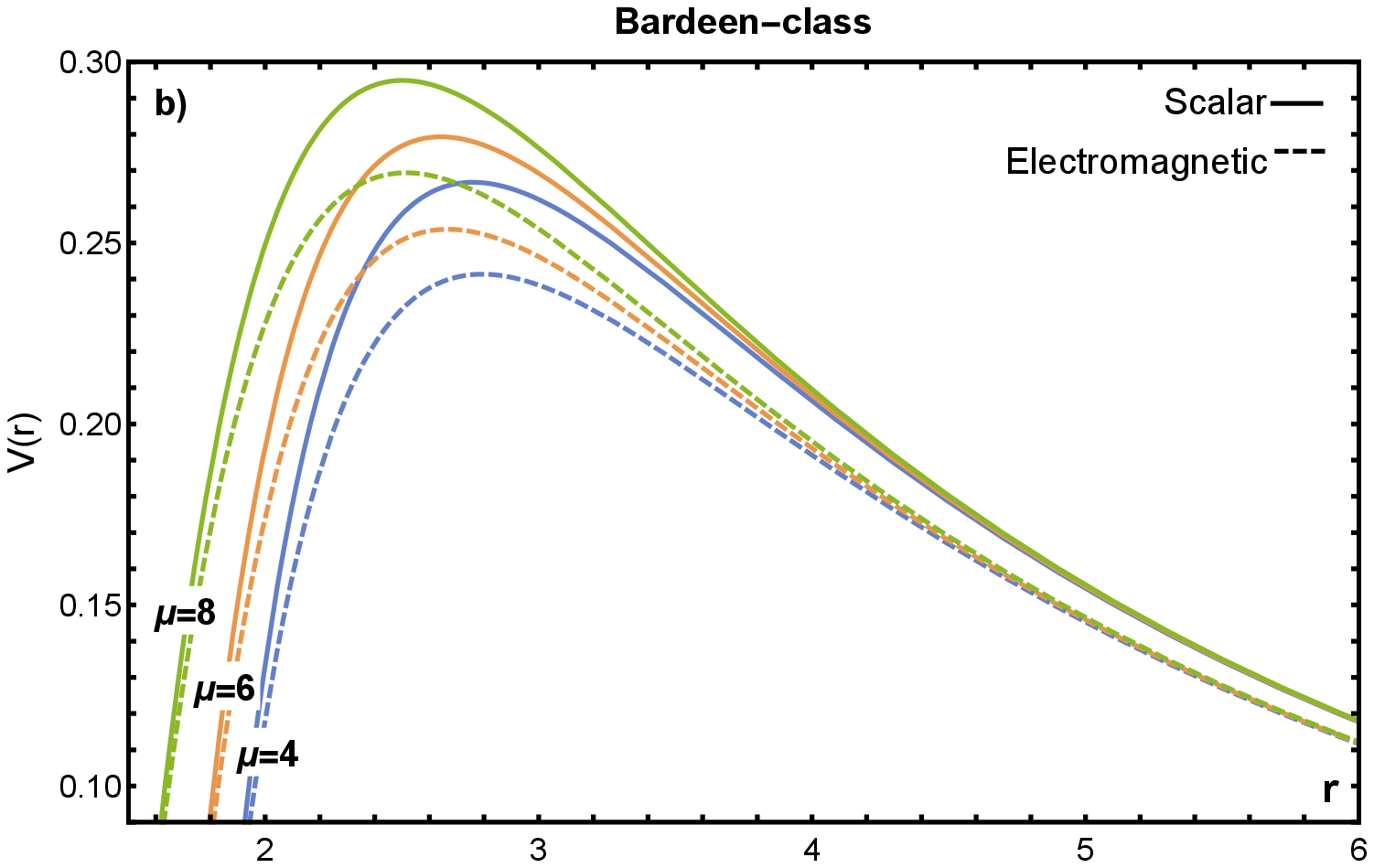}
\includegraphics [width =0.48 \textwidth ]{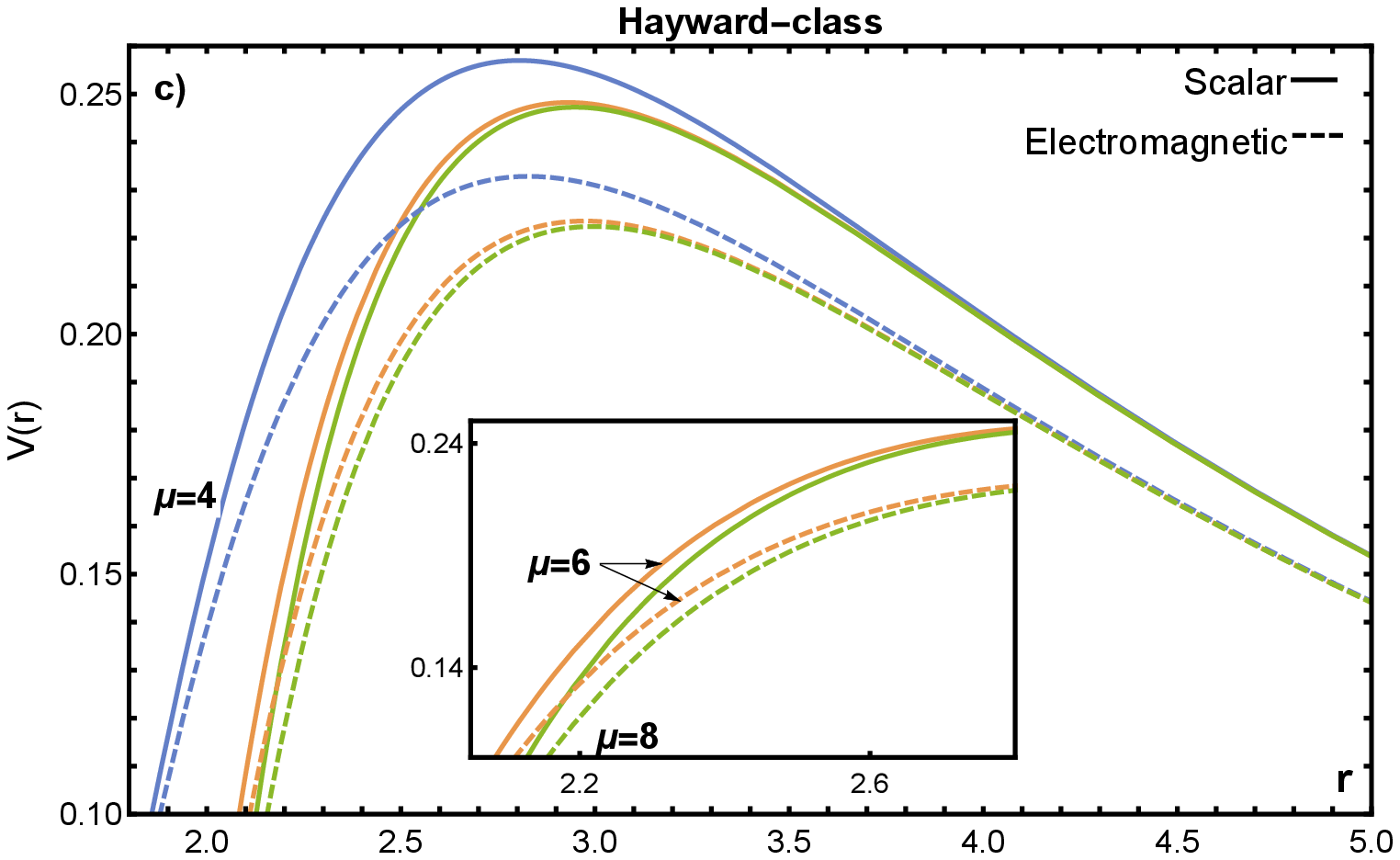}
\includegraphics [width =0.48 \textwidth ]{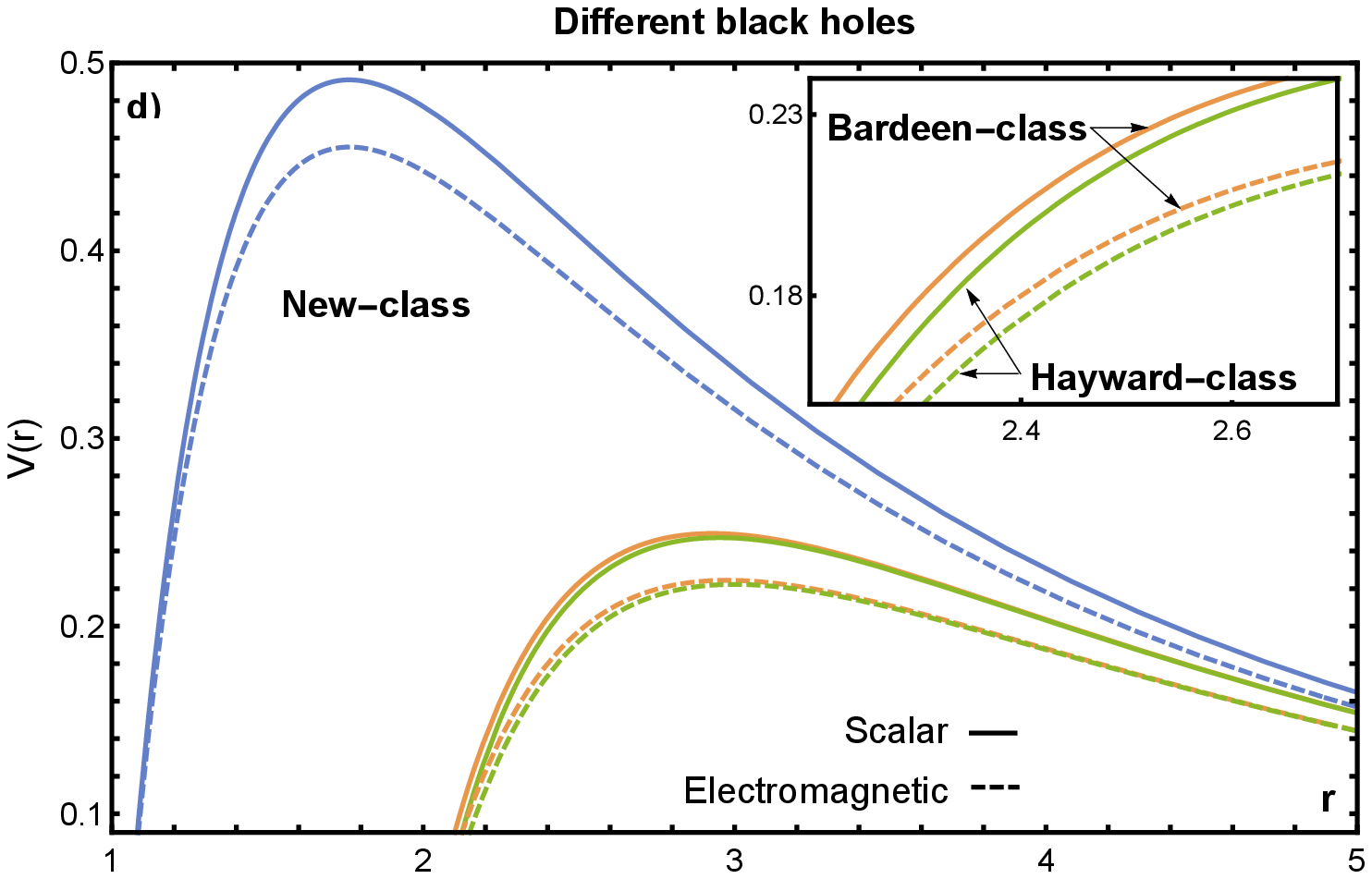}
\end{center}
\caption{ a) The effective potential for the Newc is shown for various values of $\mu$, with $q=0.07$. b) The effective potential for the Barc is shown for various values of $\mu$, with $q=0.41$. c) The effective potential for the Hawc is shown for various values of $\mu$, with $q=1.13$. d) The effective potential for the different BH is compared with $\mu=6$,  $q=0.12$. In the figures, the solid line denotes the effective potential of the scalar perturbation and the dashed line the  electromagnetic perturbation. We fix the all cases the values of $M=1$ and $l=2$.}\label{Fig2}
\end{figure}

In the Fig \ref{Fig2}, we consider the value of $l=2$ and the dependence of effective potentials is shown (\ref{ec.poseg}) with the parameter $\mu$ for Newc (Fig \ref{Fig2}  a)), Barc (Fig \ref{Fig2}  b)) and Hawc (Fig \ref{Fig2}  c)). It can be noticed that in the case of Newc and Barc the potentials increase as $\mu$ increases. When $\mu$ is increased, the maximum of potentials increase and their position is shifted to the left for both perturbations, this implies that the strength of nonlinearity of the electromagnetic field increases the magnitudes of the different potentials. But in the case of Hawc, when $\mu$ increases, the  maximum of potentials decrease and its position is shifted to the right. It is possible to observe that $V(r)_{elec}< V(r)_{sc}$ for the three different cases of black holes.

On the other hand, in Fig. \ref{Fig2} d) the different effective potentials of the three BHs are compared with the same charge. We can observe that $V(r)_{Hawc} \lesssim V(r)_{RN}<V(r)_{Barc}<V(r)_{Newc}$ in both perturbations, we note that the scalar and electromagnetic perturbations of the Newc are higher than the scalar and  electromagnetic perturbations of Hawc and Barc. The difference between Barc and Hawc potentials is not significant as is shown in the Fig \ref{Fig2} d) . To complete the comparison between the effective potentials, we introduce $V(r)_{RN}$ that represents the effective potential of a Reissner-Nordstrom (RN) BH. In the following sections all the results are compared with the RN BH which is the linear counterpart of nonlinear electromagnetic BHs.

Lastly, we can mention that electromagnetic and scalar perturbations have similar behavior. However, the maximum potential height is much higher for scalar perturbations, and the maximum of the potential depends on $\mu$. It is also possible to mention that the maximum height of the potentials increases as $l$ increases with a $\mu$ fixed. The effective potential $V(r)$ has asymptotic value $V(r) \approx 0$ when $r\to\infty$, in both cases, scalar and  electromagnetic perturbation.

\section{Quasinormal modes of Black Holes}

The QNM are obtained from the solutions of the wave equation (\ref{ecmcnsf}),  in the event horizon the waves have to be purely ingoing $\tilde{\xi}(r)\sim e^{-i\omega r_*}$ and purely outgoing at spatial infinity $\tilde{\xi}(r)\sim e^{i\omega r_*}$, the  wave function $\Phi$ has a time dependency as  $\Phi(t,r,\phi,\theta)\sim e^{-i\omega t}$. There exists a discrete set of complex frequencies satisfying these conditions: $\omega=\omega_r-i|\omega_i|$.

The real part of the frequency, $\omega_r$ represents the oscillation of the black hole, which is always positive. On the other hand, the imaginary part of the frequency $\omega_i$ can be positive or negative. In the case where $\omega_i<0$, we have a damped oscillation, because $\Phi$ decreases as time progresses and eventually tends to zero for very long periods of time, for which the oscillation of the black hole will gradually cease (stable solution).

Several methods are proposed to calculate the quasinormal modes, but the WKB method is one of the most used. Using the code provided in \cite{Konoplya:2019hlu}, in the tables \ref{Tab1} and \ref{Tab2} we show the calculated frequencies with different orders of WKB and different values of $\mu$.

In Table \ref{Tab1}, the spectrum for the $\omega_{r}$ frequencies of the QNM of the scalar and electromagnetic perturbations is shown up to the sixth order. We can observe that in all orders, the  $\omega_{r_{sc}}$ and $\omega_{r_{elec}}$ of the Newc are more significant than the frequencies of Barc and Hawc. On the other hand, the values of $\omega_{r}$ of Barc are very close to the values of $\omega_{r}$ of Hawc for both perturbations.

The $|\omega_{i}|$ of the scalar and electromagnetic perturbations are shown in the Table \ref{Tab2}, we can observe that in all orders  the values of $|\omega_{i}|$ are very close for the three--class black holes and in general $ |\omega_{i_{sc}}|  > | \omega_{i_{elec}}|$.

\begin{table}{\tiny
\begin{ruledtabular}
\begin{tabular}{llllllllll}
 $\mu$ & q & WKB & \multicolumn{3}{c}{Scalar}  & \multicolumn{3}{c}{Electromagnetic}    \\ 
\cline{4-6} \cline{7-9} 
     &   &  order   & New--class & Bardeen--class           
	&  Hayward--class   & New--class & Bardeen--class & Hayward--class                                                  \\    \hline
 \multirow{6}{*}{4} & \multirow{6}{*}{0.17} & 1 & 0.729515 & 0.563434 & 0.561097 & 0.703442 & 0.537951 & 0.535532\\ 
                                          &  & 2 & 0.656427 & 0.475397 & 0.471966 & 0.629307 & 0.449102 & 0.445595\\
                                          &  & 3 & 0.652737 & 0.466936 & 0.463196 & 0.625257 & 0.439689 & 0.435833\\
                                          &  & 4 & 0.653249 & 0.467554 & 0.463759 & 0.625848 & 0.440358 & 0.436435\\
                                          &  & 5 & 0.653299 & 0.467683 & 0.463866 & 0.625916 & 0.440495 & 0.436541\\
                                          &  & 6 & 0.653298 & 0.467647 & 0.463853 & 0.625903 & 0.440464 & 0.436542\\ \hline
 \multirow{6}{*}{6} & \multirow{6}{*}{0.12} & 1 & 0.751093 & 0.562837 & 0.561096 & 0.725453 & 0.537333 & 0.535531\\ 
                                          &  & 2 & 0.682926 & 0.474519 & 0.471961 & 0.656535 & 0.448203 & 0.445589\\
                                          &  & 3 & 0.679575 & 0.46598 & 0.463192 & 0.652914 & 0.438702 & 0.435828\\
                                          &  & 4 & 0.680296 & 0.466583 & 0.463753 & 0.653722 & 0.439353 & 0.436428\\
                                          &  & 5 & 0.680348 & 0.466706 & 0.463859 & 0.653791 & 0.439483 & 0.436533\\
                                          &  & 6 & 0.680345 & 0.466677 & 0.463847 & 0.653776 & 0.43946 & 0.436534\\ \hline
 \multirow{6}{*}{8} & \multirow{6}{*}{0.07} & 1 & 0.687894 & 0.561882 & 0.561096 & 0.661698 & 0.536344 & 0.535531\\ 
                                          &  & 2 & 0.609731 & 0.473113 & 0.471961 & 0.582411 & 0.446766 & 0.445589\\
                                          &  & 3 & 0.605126 & 0.464448 & 0.463192 & 0.577315 & 0.437123 & 0.435828\\
                                          &  & 4 & 0.605639 & 0.465028 & 0.463753 & 0.577918 & 0.437745 & 0.436428\\
                                          &  & 5 & 0.605711 & 0.465142 & 0.463859 & 0.578015 & 0.437862 & 0.436533\\
                                          &  & 6 & 0.605696 & 0.465122 & 0.463847 & 0.577981 & 0.437852 & 0.436534\\
\end{tabular}
\end{ruledtabular}}\caption {Quasi--normal frequencies $\omega_{r}$  for scalar and electromagnetic perturbation for different order WKB and different values of $\mu$, $l=2$, $n=0$ and $M=1$.} 
\label{Tab1}
\end{table}  

\begin{table}{\tiny
\begin{ruledtabular}
\begin{tabular}{llllllllll}
 $\mu$ & q & WKB & \multicolumn{3}{c}{Scalar}  & \multicolumn{3}{c}{Electromagnetic}    \\ 
\cline{4-6} \cline{7-9} 
     &   &  order   & New--class & Bardeen--class           
	&  Hayward--class  & New--class & Bardeen--class & Hayward--class                                                 \\    \hline
  \multirow{6}{*}{4} & \multirow{6}{*}{0.17} & 1 & 0.277043 & 0.259483 & 0.260209 & 0.273083 & 0.253476 & 0.254101\\ 
    &  & 2 & 0.307889 & 0.307535 & 0.309349 & 0.305254 & 0.303623 & 0.305387\\
    &  & 3 & 0.299943 & 0.294286 & 0.295797 & 0.296814 & 0.289519 & 0.290958\\
    &  & 4 & 0.299708 & 0.293898 & 0.295438 & 0.296534 & 0.289079 & 0.290557\\
    &  & 5 & 0.299817 & 0.294103 & 0.295606 & 0.296678 & 0.289289 & 0.290715\\
    &  & 6 & 0.299818 & 0.294125 & 0.295614 & 0.296684 & 0.289309 & 0.290715 \\ \hline
 \multirow{6}{*}{6} & \multirow{6}{*}{0.12} & 1 & 0.270338 & 0.259675 & 0.260216 & 0.266525 & 0.253643 & 0.254108\\ 
     &  & 2 & 0.297321 & 0.308007 & 0.309361 & 0.294503 & 0.304083 & 0.305399\\
     &  & 3 & 0.289541 & 0.294682 & 0.29581 & 0.286339 & 0.289896 & 0.290972\\
     &  & 4 & 0.289235 & 0.294301 & 0.295452 & 0.285984 & 0.289467 & 0.290571\\
     &  & 5 & 0.289357 & 0.294496 & 0.295619 & 0.286141 & 0.289663 & 0.290729 \\
     &  & 6 & 0.289358 & 0.294515 & 0.295627 & 0.286147 & 0.289678 & 0.290728\\ \hline
 \multirow{6}{*}{8} & \multirow{6}{*}{0.07} & 1 & 0.27655 & 0.259977 & 0.260216 & 0.272104 & 0.253903 & 0.254108\\ 
     &  & 2 & 0.312002 & 0.308756 & 0.309361 & 0.309147 & 0.304812 & 0.305399\\
     &  & 3 & 0.302904 & 0.295303 & 0.29581 & 0.299436 & 0.290493 & 0.290972\\
     &  & 4 & 0.302647 & 0.29494 & 0.295452 & 0.299124 & 0.29008 & 0.290571\\
     &  & 5 & 0.30279 & 0.29512 & 0.295619 & 0.299311 & 0.290255 & 0.290729\\
     &  & 6 & 0.302798 & 0.295133 & 0.295627 & 0.299329 & 0.290262 & 0.290728\\
\end{tabular}
\end{ruledtabular}}\caption {Quasi--normal frequencies $|\omega_{i}|$  for scalar and electromagnetic perturbation for different order WKB and different values of $\mu$, $l=2$, $n=0$ and $M=1$.} 
\label{Tab2}
\end{table}
    
This paper also uses the third--order WKB approximation developed by Schutz, Will \cite{Schutz:1985km}, and Iyer \cite{Iyer:1986np}. The expression for obtaining the quasinormal frequencies is given by:

\begin{equation}\label{ec.fmcn1}
\omega^2=\left[V_{max}+\left(-2V^{(2)}_{max}\right)^{1/2}\Lambda(\alpha)\right]-i \alpha\left(-2V^{(2)}_{max}\right)^{1/2}\left[1+\Omega(\alpha)\right]\,,
\end{equation}
where

{\small
\begin{eqnarray}
\Lambda (\alpha)&=&\frac{1}{\left(-2V^{(2)}_{max}\right)^{1/2}}\left\{
\frac{1}{8}\left(\frac{V_{max}^{(4)}}{V^{(2)}_{max}}\right)\left(\frac{1}{4}+\alpha^2\right)
-\frac{1}{288}\left(\frac{{V^{(3)}}_{max}}{V^{(2)}_{max}}\right)^2\left(7+60\alpha^2\right)
\right\}\,,\label{ecu.wkbl}\\
\Omega (\alpha)&=&\frac{1}{-2V^{(2)}_{max}}
\left\{
\frac{5}{6912}\left(\frac{V^{(3)}_{max}}{V^{(2)}_{max}}\right)^4\left(77+188\alpha^2\right)
-
\frac{1}{384}\left(\frac{{V^{(3)}}_{max}^2V_{max}^{(4)}}{{V^{(2)}}_{max}^3}\right)\left(51+100\alpha^2\right)
\right.\nonumber\\
&&
-\left.
\frac{1}{288}\left(\frac{V_{max}^{(6)}}{V^{(2)}_{max}}\right)\left(5+4\alpha^2\right)
+
\frac{1}{288}\left(\frac{{V^{(3)}}_{max}V_{max}^{(5)}}{{V^{(2)}}_{max}^2}\right)\left(19+28\alpha^2\right)
+
\frac{1}{2304}\left(\frac{V_{max}^{(4)}}{V^{(2)}_{max}}\right)^2\left(67+68\alpha^2\right)
\right\}\label{ecu.wkbo}\,,
\end{eqnarray}}
with
\begin{equation}
\alpha=n+\frac{1}{2}
,\quad V_{max}^{(m)}=\frac{d^mV(r)}{dr_*^m}\Big|_{r_*=r_*(r_q)}
\,,
\end{equation}
where $r_*(r_q)$ indicates the value of  $r_*$ at which the effective potential ($V(r)$) obtains its maximum ($V_{max}$) and $n$ is the oscillation mode.  

The WKB method works best for $l>n$, while for $l=n$ it is not appropriate, as other authors have shown (see \cite{Konoplya:2003ii}). 
The third--order WKB approximation method is used and the Tables \ref{Tab2.1}--\ref{Tab2.2} show the spectrum of the frequencies ($\omega_{r}$ and $|\omega_{i}|$) of the QNM for scalar and  electromagnetic  perturbations with different values of $n$, $l$ and $q$. 

It can be seen that the range of real frequencies is approximately the same for each case of $l$ and it does not change drastically, despite its value of $n$, however, an increase in $n$ for the same $l$ will lead to a decrease in the frequencies (see Table  \ref{Tab2.1}).

In contrast, the range of imaginary frequencies is approximately the same for each case of $n$ and it does not change drastically, despite its value of $l$. We can also observe that for each value of $n$ and $l$, the highest value of $|\omega_{i}|$ for the Newc is when we take the middle value of the range of charge that was chosen ($q=0.08$), for Barc the highest value of $|\omega_{i}|$ is when we take the smallest value of the range of charge that was chosen ($q=0.28$), for Hawc the same behavior is followed ($q=0.01$), but the differences between frequencies is much smaller for this case (see Table  \ref{Tab2.2}). 

\squeezetable
\begin{table} {\tiny
\begin{ruledtabular}
\begin{tabular}{lllllllllll}
 n & l & \multicolumn{3}{c}{New-class} & \multicolumn{3}{c}{Bardeen-class}  & \multicolumn{3}{c}{Hayward-class}   \\ 
\cline{3-5} \cline{6-8} \cline{9-11} 
     &   &  $q$  & $\omega_{r_{sc}}$ & $\omega_{r_{elec}}$           
	&  $q$   & $\omega_{r_{sc}}$ & $\omega_{r_{elec}}$ & $q$ & $\omega_{r_{sc}}$ & $\omega_{r_{elec}}$                                                   \\    \hline
 \multirow{9}{*}{0}  & \multirow{3}{*}{2} & 0.04 & 0.528454 & 0.501807 & 0.10 & 0.485478 & 0.459508 & 0.01 & 0.483826 & 0.45782 \\
                                          &  & 0.08 & 0.590492 & 0.563444 & 0.28 & 0.497737 & 0.472085 & 0.51 & 0.483831 & 0.457828 \\
                                          &  & 0.12 & 0.691364 & 0.665145 & 0.46 & 0.528462 & 0.504077 & 1.01 & 0.484073 & 0.458247\\ \cline{2-11}
                     & \multirow{3}{*}{3} & 0.04 & 0.73759 & 0.718762 & 0.10 & 0.677706 & 0.659271 & 0.01 & 0.675412 & 0.656952\\
                                          &  & 0.08 & 0.824304 & 0.805168 & 0.28 & 0.694764 & 0.67654 & 0.51 & 0.675424 & 0.656966 \\
                                          &  & 0.12 & 0.965995 & 0.947399 & 0.46 & 0.737979 & 0.720604 & 1.01 & 0.676108 & 0.657755 \\ \cline{2-11}
                    & \multirow{3}{*}{4} & 0.04 & 0.94736 & 0.932684 & 0.10 & 0.870365 & 0.856065 & 0.01 & 0.867428 & 0.853109\\
                                          &  & 0.08 & 1.05873 & 1.04382 & 0.28 & 0.892222 & 0.878079 & 0.51 & 0.867447 & 0.853125 \\
                                          &  & 0.12 & 1.24107 & 1.22659 & 0.46 & 0.947831 & 0.934328 & 1.01 & 0.868486 &  0.854243\\ \hline
 \multirow{9}{*}{1} & \multirow{3}{*}{3} & 0.04 & 0.72641 & 0.707212 & 0.10 & 0.665529 & 0.646809 & 0.01 & 0.66308 & 0.6443311 \\
                                          &  & 0.08 & 0.814804 & 0.79535 & 0.28 & 0.683676 & 0.66521 & 0.51 & 0.663081 & 0.644335\\
                                          &  & 0.12 & 0.958536 & 0.939784 & 0.46 & 0.728159 & 0.710708 & 1.01 & 0.663006 & 0.644425 \\ \cline{2-11}
                     & \multirow{3}{*}{4} & 0.04 & 0.937807 & 0.922973 & 0.10 & 0.859952 & 0.845487 & 0.01 & 0.856873 & 0.842386 \\
                                          &  & 0.08 & 1.05072 & 1.03568 & 0.28 & 0.882803 & 0.868523 & 0.51 & 0.856883 & 0.842398 \\
                                          &  & 0.12 & 1.23496 & 1.22042 & 0.46 & 0.939711 & 0.926164 & 1.01 & 0.857355 & 0.842975 \\ \cline{2-11}
                     & \multirow{3}{*}{5} & 0.04 & 1.14908 & 1.13699 & 0.10 & 1.05427 & 1.04249 & 0.01 & 1.05057 & 0.03877 \\
                                          &  & 0.08 & 1.28648 & 1.27422 & 0.28 & 1.08181 & 1.07017 & 0.51 & 1.05058 & 1.03878  \\
                                          &  & 0.12 & 1.51114 & 1.49927 & 0.46 & 1.151 & 1.13993 & 1.01 & 1.0515 & 1.03977 \\ \hline
 \multirow{9}{*}{2} & \multirow{3}{*}{4} & 0.04 & 0.921765 & 0.906733 & 0.10 & 0.842871 & 0.828218 & 0.01 & 0.839592 & 0.824914 \\
                                          &  & 0.08 & 1.03686 & 1.02161 & 0.28 & 0.867128 & 0.852687 & 0.51 & 0.839585 & 0.824911\\
                                          &  & 0.12 & 1.22378 & 1.20916 & 0.46 & 0.925534 & 0.911952 & 1.01 & 0.838932 & 0.824414 \\ \cline{2-11}
                     & \multirow{3}{*}{5} & 0.04 & 1.13452 & 1.12229 & 0.10 & 1.03858 & 1.02666 & 0.01 & 1.03468 & 1.02273 \\
                                          &  & 0.08 & 1.27409 & 1.26171 & 0.28 & 1.06754 & 1.05578 & 0.51 & 1.03468 & 1.02274 \\
                                          &  & 0.12 & 1.50148 & 1.48955 & 0.46 & 1.13845 & 1.12735 & 1.01 & 1.03466 & 1.02283\\ \cline{2-11}
                     & \multirow{3}{*}{6} & 0.04 & 1.34706 & 1.33677 & 0.10 & 1.23418 & 1.22414 & 0.01 & 1.22966 & 1.2196 \\
                                          &  & 0.08 & 1.51106 & 1.50062 & 0.28 & 1.26777 & 1.25786 & 0.51 & 1.22967 & 1.21961 \\
                                          &  & 0.12 & 1.77871 & 1.76864 & 0.46 & 1.35095 & 1.34156 & 1.01 & 1.23019 & 1.22021 \\ 
\end{tabular}
\end{ruledtabular}}
\caption {Third--order WKB approximation of $\omega_{r}$ for different  $n$ and $l$. The mass parameter is fixed to $M=1$ and $\mu=6$} \label{Tab2.1}
\end{table}

\squeezetable
\begin{table} {\tiny
\begin{ruledtabular}
\begin{tabular}{lllllllllll}
 n & l & \multicolumn{3}{c}{New-class} & \multicolumn{3}{c}{Bardeen-class}  & \multicolumn{3}{c}{Hayward-class}   \\ 
\cline{3-5} \cline{6-8} \cline{9-11} 
     &   &  $q$  & $|\omega_{i_{sc}}|$ & $|\omega_{i_{elec}}|$           
	&  $q$   & $|\omega_{i_{sc}}|$ & $|\omega_{i_{elec}}|$ & $q$ & $|\omega_{i_{sc}}|$ & $|\omega_{i_{ele}}|$                                                   \\    \hline
 \multirow{9}{*}{0}  & \multirow{3}{*}{2} & 0.04 & 0.101785 & 0.0994411 & 0.10 & 0.0995675 & 0.0980747 & 0.01 & 0.0998303 & 0.0983257 \\
                                          &  & 0.08 & 0.102561 & 0.101414 & 0.28 & 0.0973874 & 0.0959668 & 0.51 & 0.0998127 & 0.0983062 \\
                                          &  & 0.12 & 0.0973284 & 0.0963248 & 0.46 & 0.0895604 & 0.0881115 & 1.01 & 0.098739 & 0.0971136\\ \cline{2-11}
                     & \multirow{3}{*}{3} & 0.04 & 0.0994573 & 0.0993839 & 0.10 & 0.0977019 & 0.0968881 & 0.01 & 0.0979419 & 0.0971222\\
                                          &  & 0.08 & 0.10057 & 0.10057 & 0.28 & 0.0956815 & 0.094905 & 0.51 & 0.0979235 & 0.0971029 \\
                                          &  & 0.12 & 0.095813 & 0.0958057 & 0.46 & 0.0881237 & 0.0873408 & 1.01 & 0.0968008 & 0.0959196 \\ \cline{2-11}
                    & \multirow{3}{*}{4} & 0.04 & 0.0994535 & 0.0989944 & 0.10 & 0.0969496 & 0.0964447 & 0.01 & 0.0971806 & 0.0966722 \\
                                          &  & 0.08 & 0.100655 & 0.100263 & 0.28 & 0.094992 & 0.0945096 & 0.51 & 0.097162 & 0.096653 \\
                                          &  & 0.12 & 0.0959546 & 0.0956223 & 0.46 & 0.087543 & 0.0870585 & 1.01 & 0.0960225 & 0.095477 \\ \hline
 \multirow{9}{*}{1} & \multirow{3}{*}{3} & 0.04 & 0.304161 & 0.302129 & 0.10 & 0.297517 & 0.297517 & 0.01 & 0.298316 & 0.296072 \\
                                          &  & 0.08 & 0.306483 & 0.304728 & 0.28 & 0.290895 & 0.28874 & 0.51 & 0.298263 & 0.296015 \\
                                          &  & 0.12 & 0.290761 & 0.289206 & 0.46 & 0.267254 & 0.264991 & 1.01 & 0.295022 & 0.292574 \\ \cline{2-11}
                     & \multirow{3}{*}{4} & 0.04 & 0.300728 & 0.299424 & 0.10 & 0.293634 & 0.292205 & 0.01 & 0.294379 & 0.292942 \\
                                          &  & 0.08 & 0.303735 & 0.302611 & 0.28 & 0.287393 & 0.286017 & 0.51 & 0.294324 & 0.292884 \\
                                          &  & 0.12 & 0.288874 & 0.287901 & 0.46 & 0.264375 & 0.262956 & 1.01 & 0.290947 & 0.289391 \\ \cline{2-11}
                     & \multirow{3}{*}{5} & 0.04 & 0.29894 & 0.29804 & 0.10 & 0.291593 & 0.290606 & 0.01 & 0.292308 & 0.291315 \\
                                          &  & 0.08 & 0.302324 & 0.30155 & 0.28 & 0.285564 & 0.284615 & 0.51 & 0.292252 & 0.291258 \\
                                          &  & 0.12 & 0.287935 & 0.287271 & 0.46 & 0.262905 & 0.261937 & 1.01 & 0.28882 & 0.28775 \\ \hline
 \multirow{9}{*}{2} & \multirow{3}{*}{4} & 0.04 & 0.507658 & 0.505658 & 0.10 & 0.496803 & 0.494616 & 0.01 & 0.498166 & 0.495968 \\
                                          &  & 0.08 & 0.5112 & 0.509461 & 0.28 & 0.485534 & 0.483405 & 0.51 & 0.498075 & 0.495874\\
                                          &  & 0.12 & 0.484461 & 0.482903 & 0.46 & 0.445604 & 0.443324 & 1.01 & 0.492627 & 0.490216 \\ \cline{2-11}
                     & \multirow{3}{*}{5} & 0.04 & 0.502776 & 0.501365 & 0.10 & 0.491297 & 0.489757 & 0.01 & 0.49258 & 0.491033 \\
                                          &  & 0.08 & 0.507305 & 0.506081 & 0.28 & 0.480589 & 0.479096 & 0.51 & 0.492487 & 0.490937 \\
                                          &  & 0.12 & 0.481847 & 0.480774 & 0.46 & 0.441572 & 0.440004 & 1.01 & 0.486826 & 0.485141 \\ \cline{2-11}
                     & \multirow{3}{*}{6} & 0.04 & 0.499782 & 0.498736 & 0.10 & 0.487875 & 0.486735 & 0.01 & 0.489106 & 0.487959 \\
                                          &  & 0.08 & 0.504962 & 0.504058 & 0.28 & 0.477544 & 0.476441 & 0.51 & 0.489011 & 0.487863 \\
                                          &  & 0.12 & 0.480337 & 0.479554 & 0.46 & 0.439168 & 0.438026 & 1.01 & 0.483248 & 0.482006 \\ \hline
\end{tabular}
\end{ruledtabular}}\caption {Third--order WKB approximation of $\omega_{i}$ with different $n$ and $l$. The mass parameter is fixed to $M=1$ and $\mu=6$} \label{Tab2.2}
\end{table}

\begin{figure}[ht]
\begin{center}
\includegraphics [width =0.42 \textwidth ]{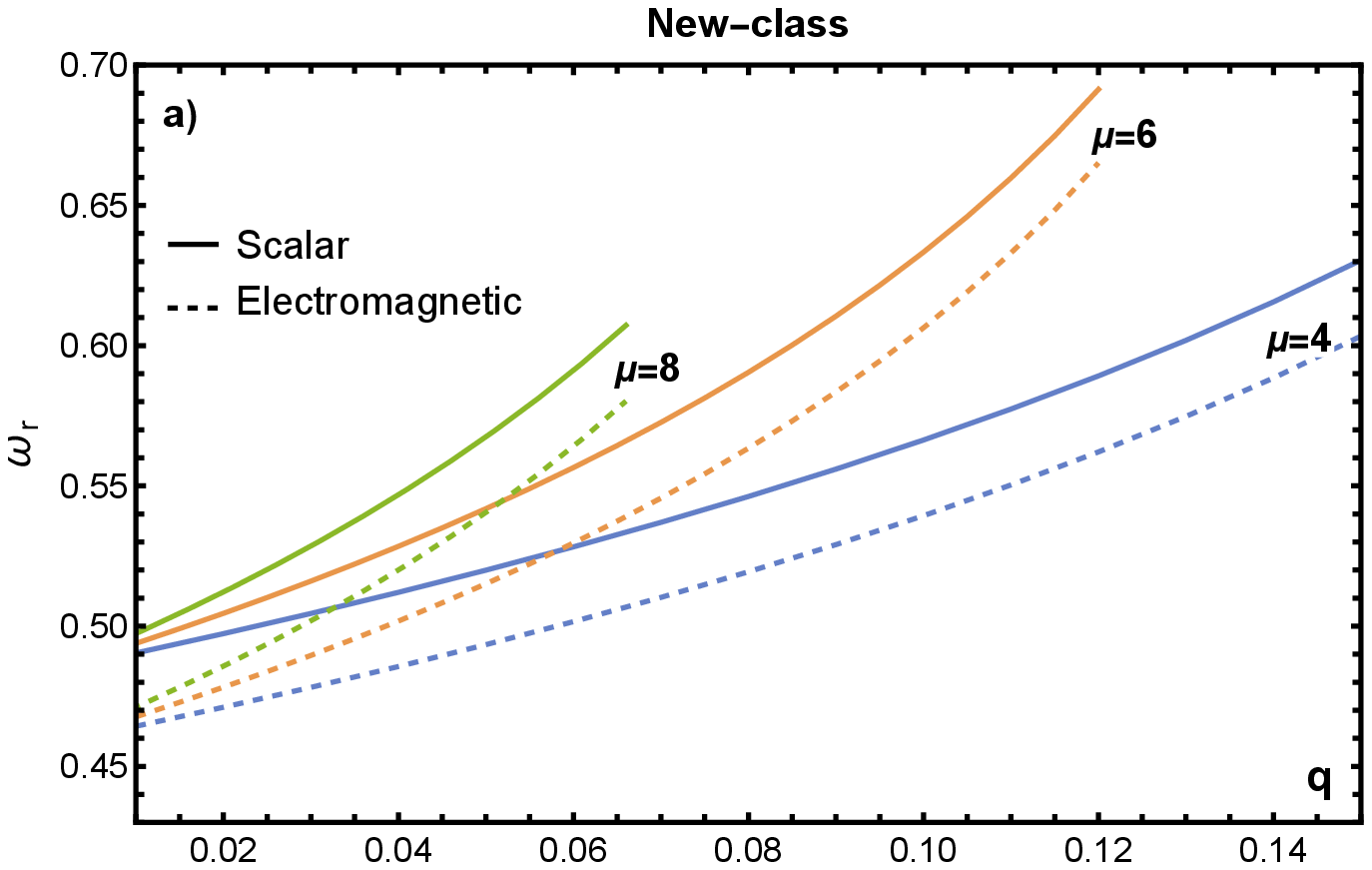}
\includegraphics [width =0.42 \textwidth ]{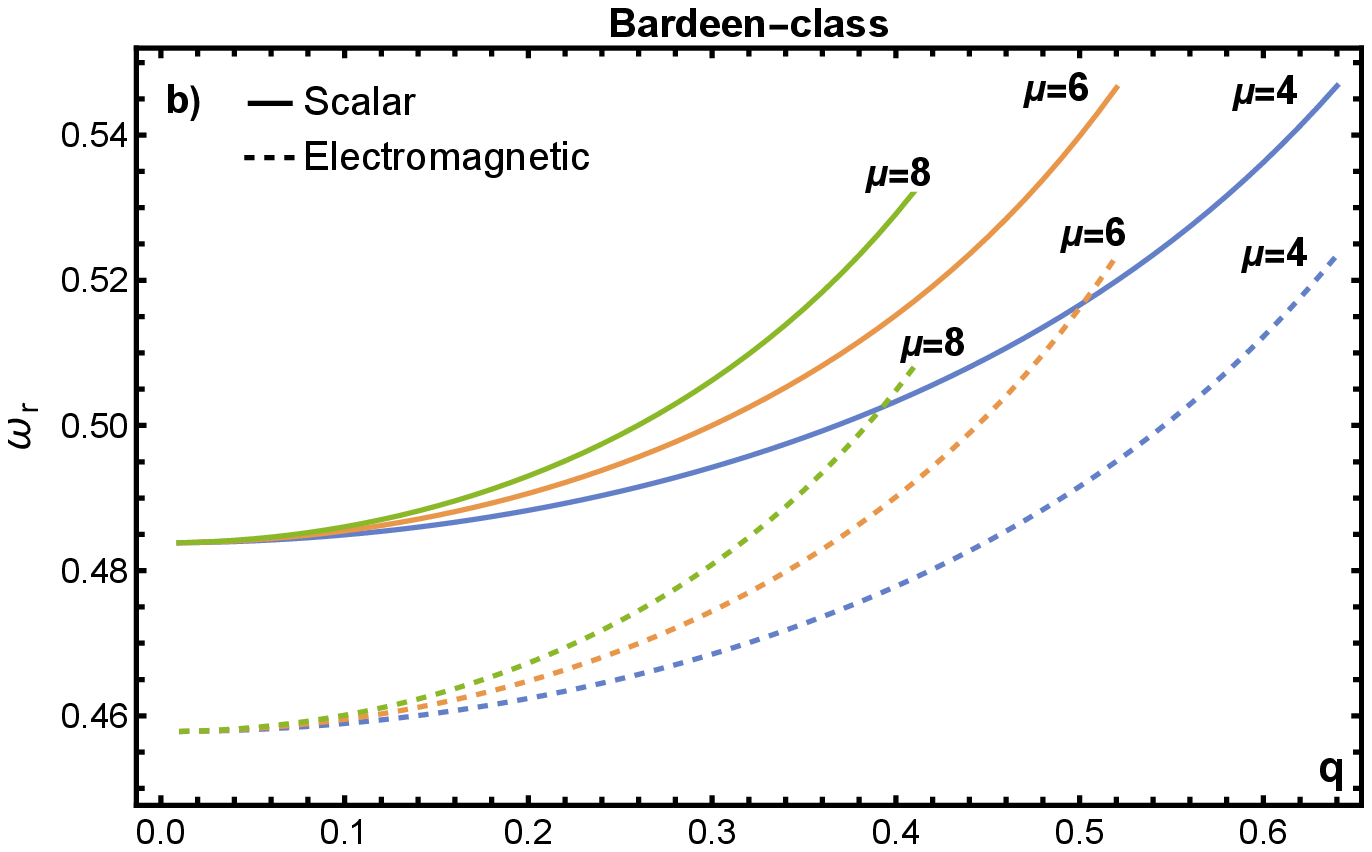}
\includegraphics [width =0.43 \textwidth ]{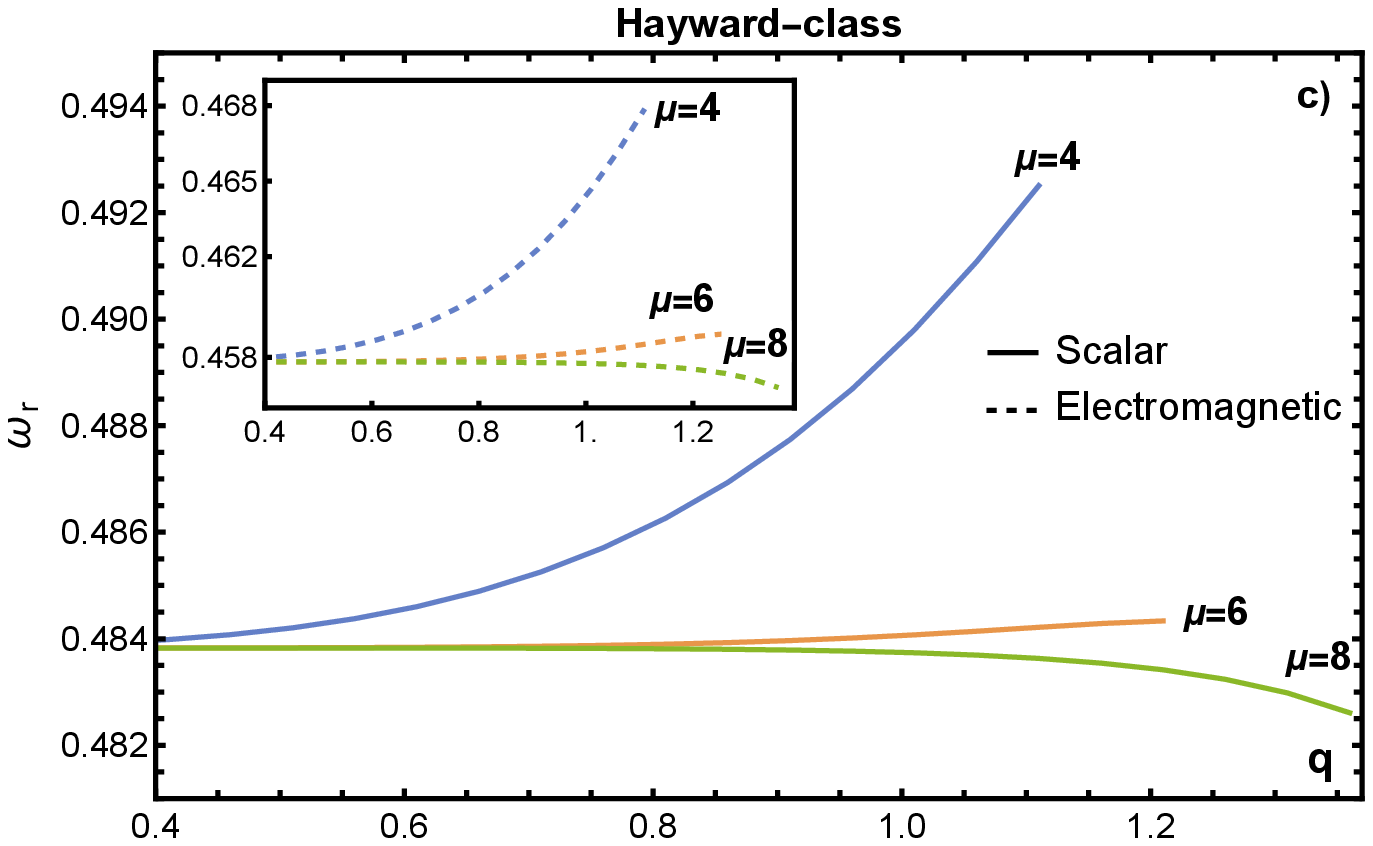}
\includegraphics [width =0.43 \textwidth ]{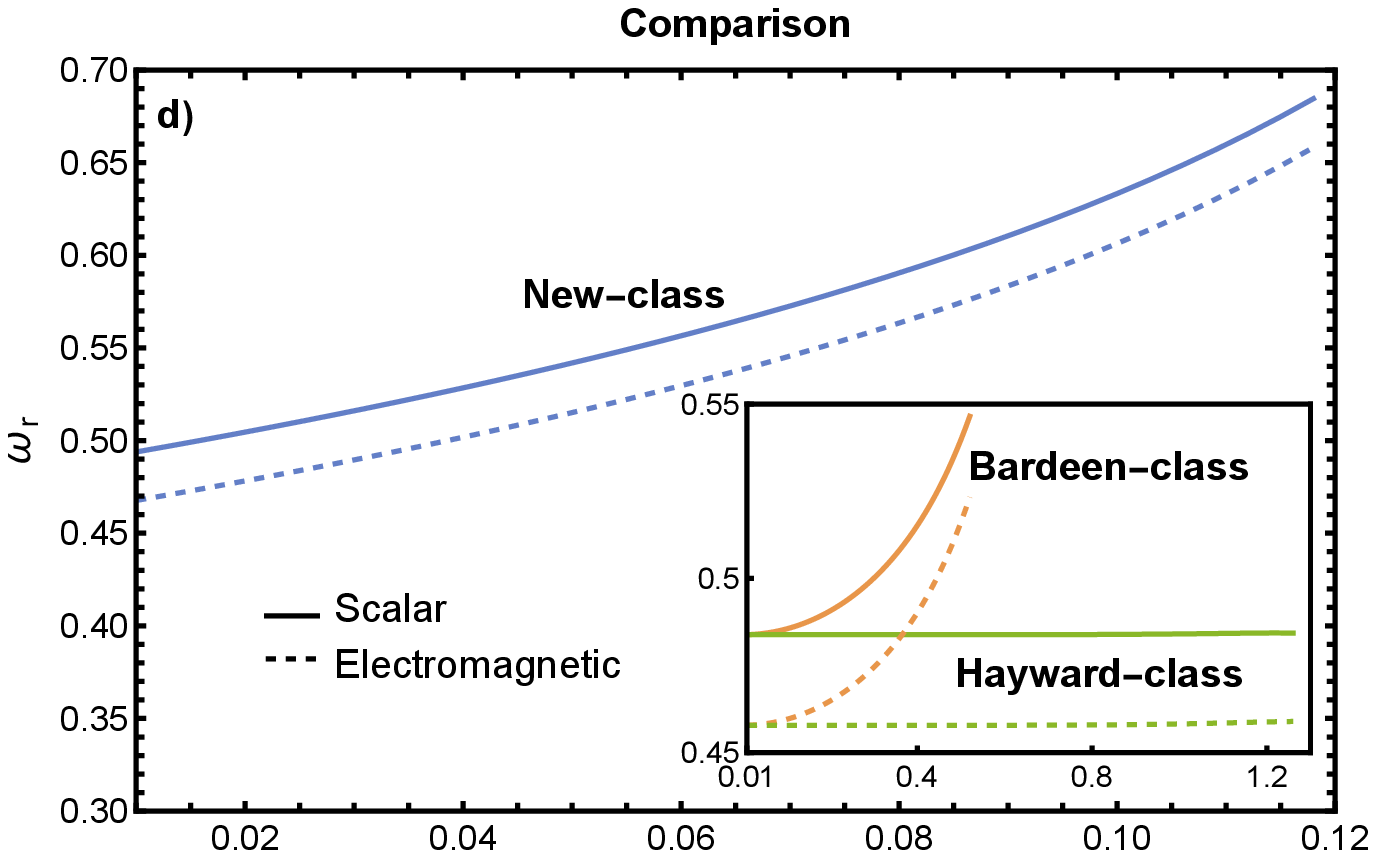}
\end{center}
\caption{The behavior of $\omega_{r}$ in the fundamental mode $n=0$ for the different BHs are shown as a function of $q$ and $l=2$. In all the figures,  the scalar perturbation is denoted with a solid line and electromagnetic perturbation is denoted with a dashed line.  a) Shows the behavior of $\omega_{r}$  for the Newc with $M=1$ and $\mu=4,6,8$. b) The behavior of $\omega_{r}$ is shown  for the Barc with $M=1$ and $\mu=4,6,8$. c) The behavior of $\omega_{r}$ is shown for the Hawc with $M=1$ and $\mu=4,6,8$. d) The behaviors of $\omega_{r}$ are compared for the different BHs with $\mu=6$ and $M=1$.}\label{Fig3}
\end{figure}

The $\omega_{r}$ frequencies as a function of magnetic charge $q$ with a fixed $n$ is shown in Fig. (\ref{Fig3}) for the scalar and electromagnetic perturbations. The Fig \ref{Fig3} a) show the dependence of $\omega_{r}$ (\ref{ec.fmcn1}) with the parameter $\mu$ for Newc, the Fig \ref{Fig3} b) for Barc and Fig \ref{Fig3} c) for Hawc. It is possible to observe that in both perturbations, the real value  ($\omega_r$)  of the QNM frequency increases when the values of $q$ increase, in the cases of Newc (Fig. \ref{Fig3} a) ) and Barc (Fig. \ref{Fig3} b)), then we can mention that the perturbation in the fundamental mode ($n = 0$) with larger values of  $q$ leads to a more intense oscillation. It is also possible to mention that $\omega_{r_{sc}}>\omega_{r_{elec}}$. Now, in both perturbations if $\mu$ increases, $\omega_{r}$ increases but the range of values for $q$ decreases i.e., the strength of nonlinearity of the electromagnetic field increases the magnitudes of oscillation and decreases the allowed $q$ values.

In the Hawc we observe (see Fig \ref{Fig3} c) )that $\omega_{r}$ increases if $\mu$ decreases and $\omega_{r_{sc}}>\omega_{r_{elec}}$, in the case $\mu=8$, $\omega_{r}$ decreases when $q$ increases, then the presence of the strength of nonlinearity of the electromagnetic field decreases the magnitudes of oscillation in the Hawc.

In the Fig \ref{Fig3} d) we show and compare the real part of the quasinormal frequencies for the three classes of BHs, in the case $\mu=6$, in general $\omega_{r_{(Newc)}}>\omega_{r_{(Barc)}} >\omega_{r_{(RN)}}>\omega_{r_{(Hawc)}}$ for both perturbations, where $\omega_{r_{(RN)}}$ represents  the real part of the quasinormal frequencies of Reissner-Nordstrom (RN) BH. Then the electromagnetic perturbations present slower oscillations than the scalar ones, the Hawc has the slowest oscillations.

\begin{figure}[h]
\begin{center}
\includegraphics [width =0.45 \textwidth ]{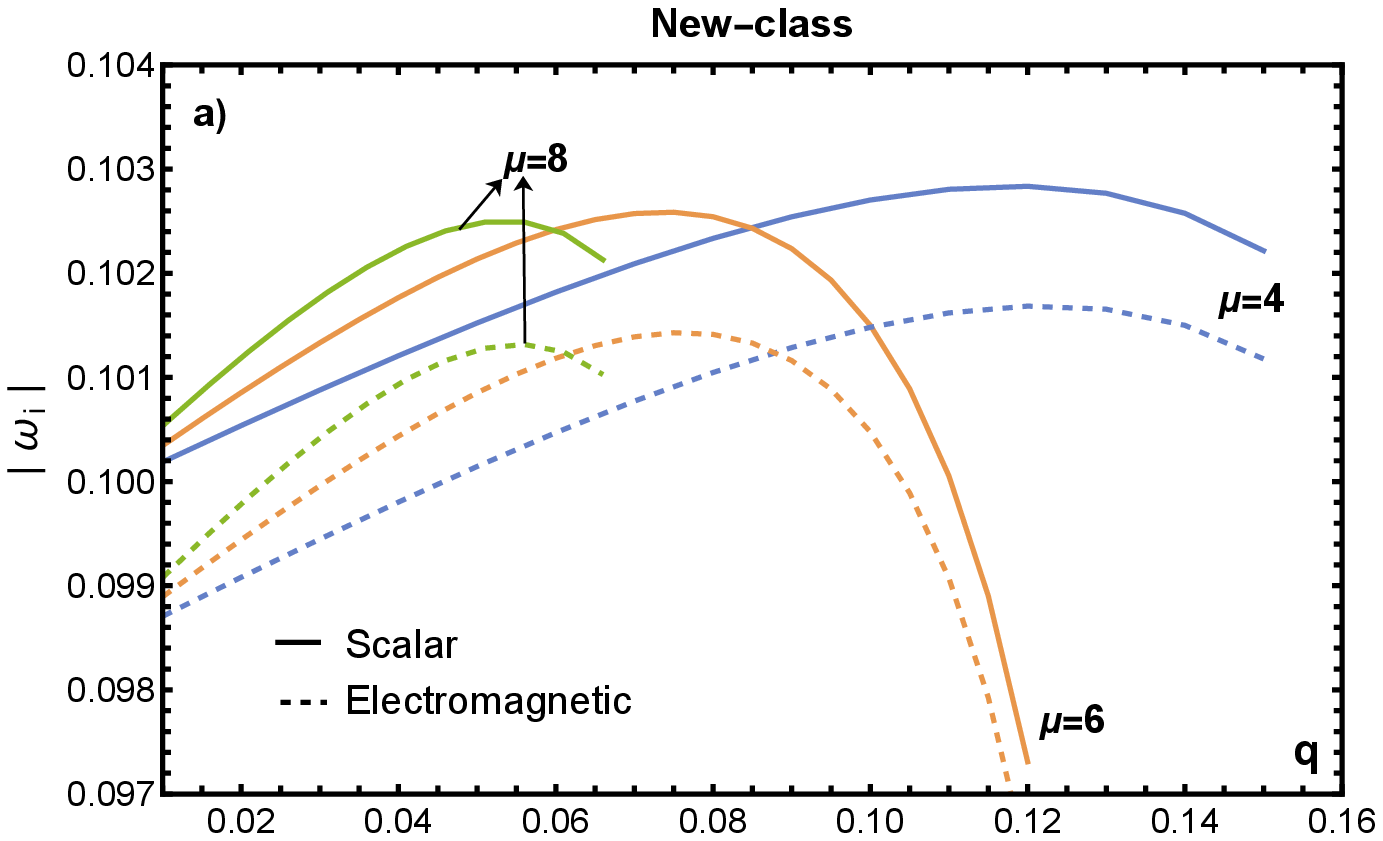}
\includegraphics [width =0.45 \textwidth ]{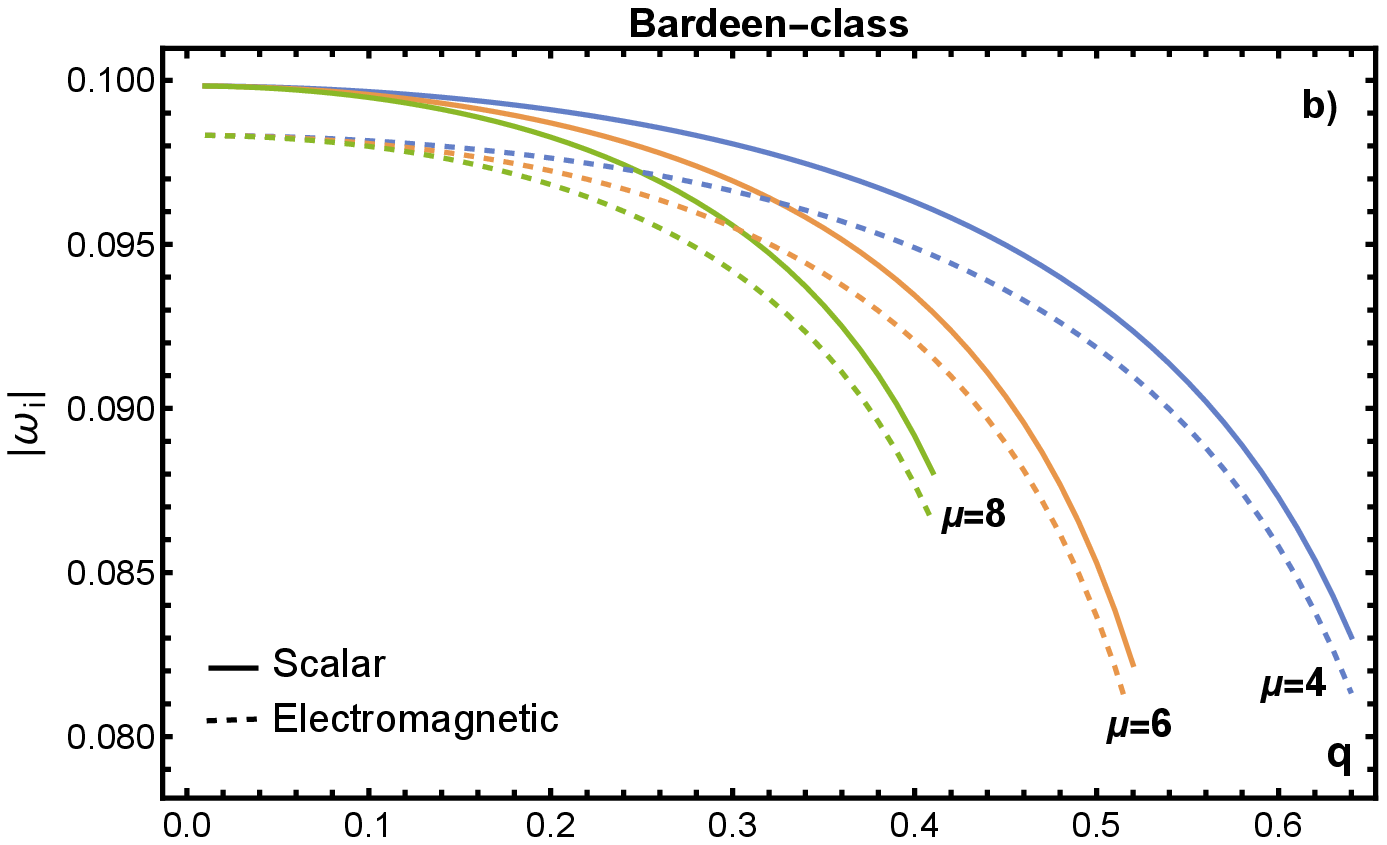}
\includegraphics [width =0.45 \textwidth ]{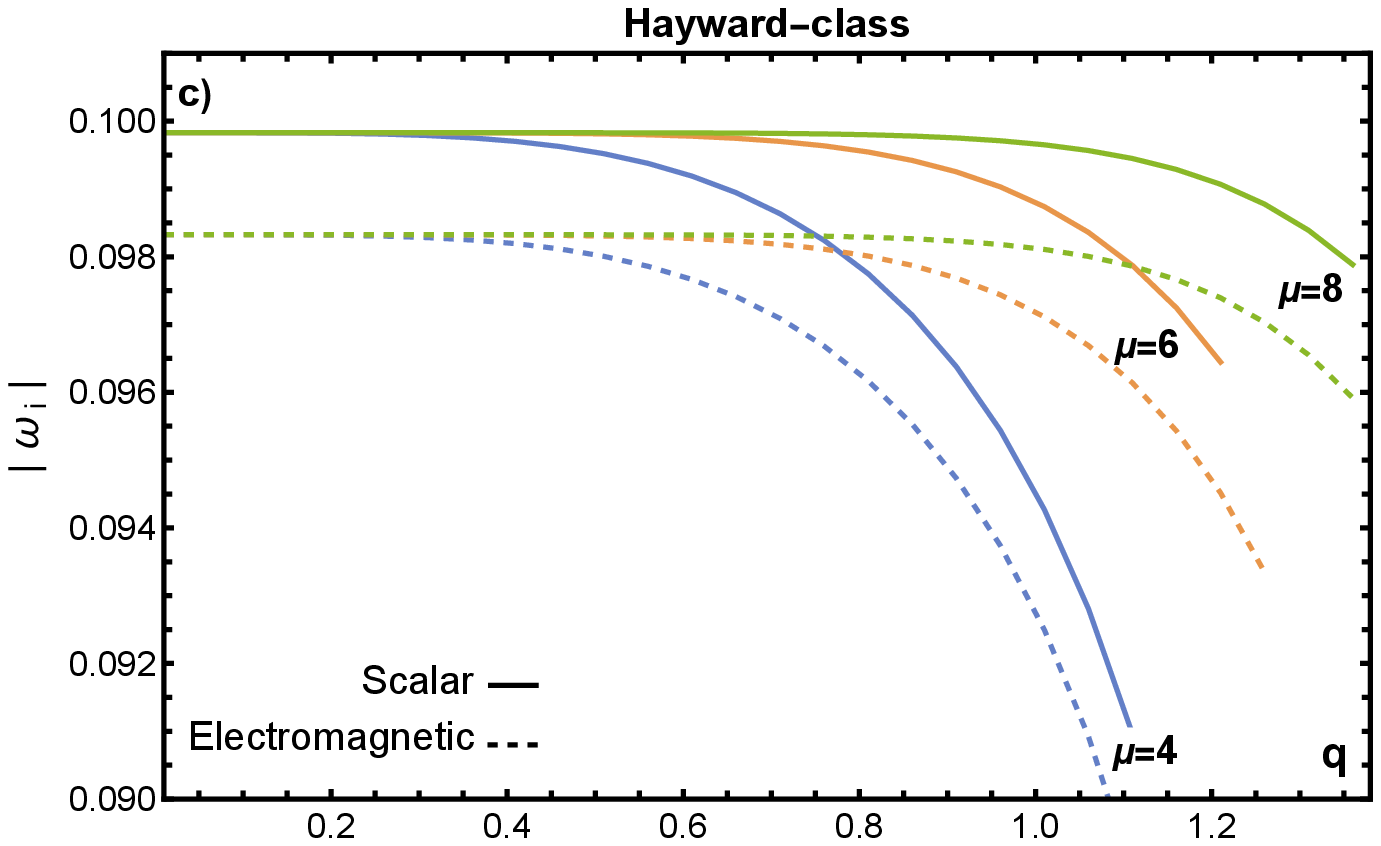}
\includegraphics [width =0.45 \textwidth ]{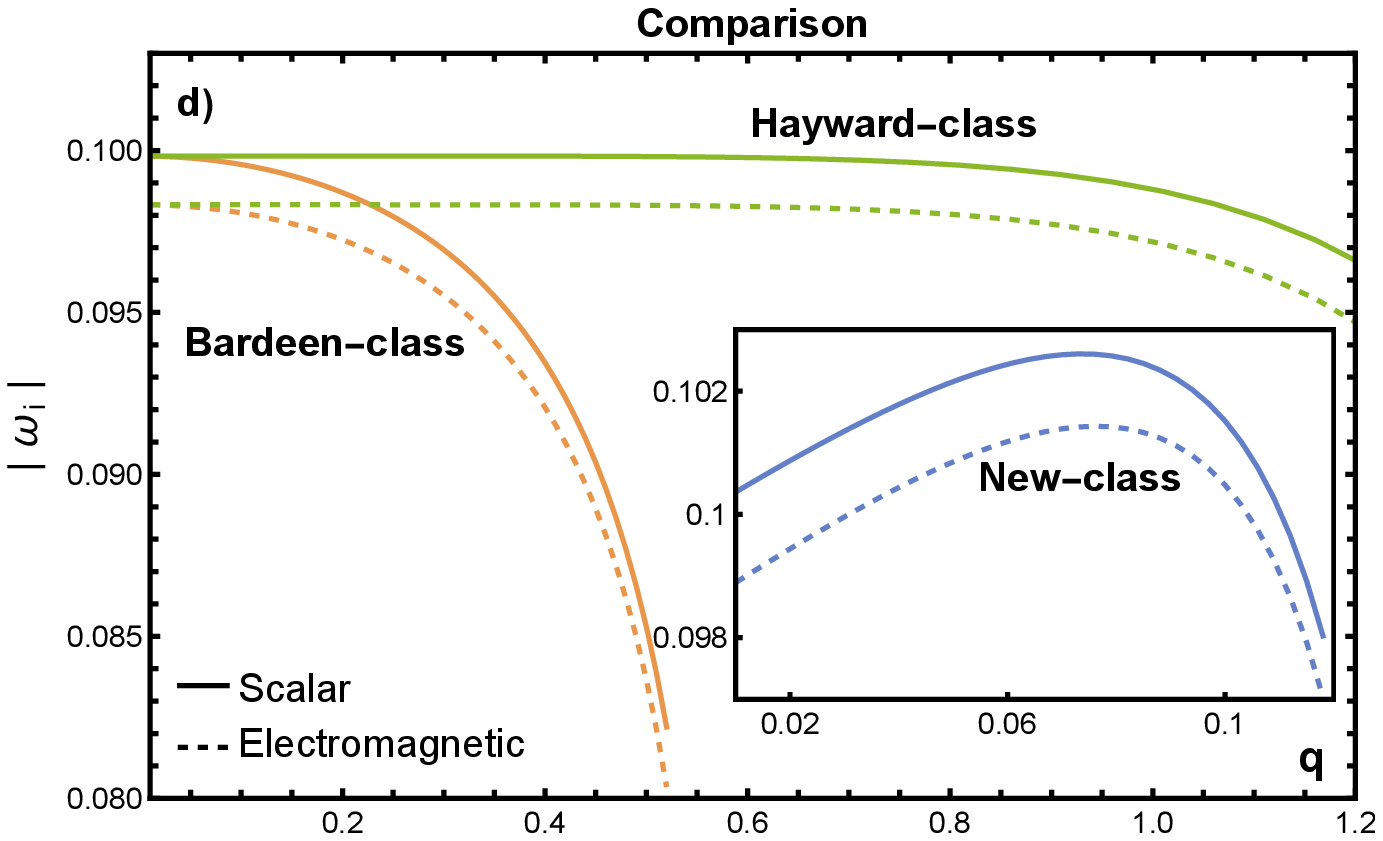}
\end{center}
\caption{The behavior of $|\omega_{i}|$ in the fundamental mode $n=0$ for the different BHs is shown as a function of $q$ and $l=2$. In all the figures, the  scalar perturbation is denoted with a solid line and electromagnetic perturbation is denoted with a dashed line.  a) Shows the behavior of $|\omega_{i}|$  for the Newc with $M=1$ and $\mu=4,6,8$. b) The behavior of $|\omega_{i}|$  is shown for the Barc with the parameter $M=1$ and $\mu=4,6,8$. c) Shows the behavior of $\omega_{i}$  for the Hawc with $M=1$ and $\mu=4,6,8$. d) The behaviors of $|\omega_{i}|$ are compared for the different BHs with $\mu=6$ and  $M=1$.}\label{Fig4}
\end{figure}

The complex quasinormal frequencies are shown in Fig. \ref{Fig4}. In the case of the Newc (see Fig. \ref{Fig4} a) ) $|\omega_i|$ increase with increasing values of $q$ and shows a maximum, then $|\omega_{i}|$ decreases. For the scalar and electromagnetic perturbations $|\omega_{i_{sc}}|>|\omega_{i_{elec}}|$. If $\mu$ increases, $\omega_{i}$ increases but the range of values for $q$ decreases.

Now for Barc and Hawc, in the Fig \ref{Fig4} b), c) we can observe that the imaginary part of QNM frequencies $|\omega_{i}|$, decreases as $q$ augments. The value of $q$ cannot exceed the value of $q_{crit}$, which corresponds to the extreme BH. Also in the Barc and Hawc $|\omega_{i_{sc}}|>|\omega_{i_{elec}}|$.

If we make the comparison of $\omega_{i}$ of the different BHs (see Fig \ref{Fig4} d)) for a fixed $\mu= 6$, we observe that in a certain range $|\omega_{i_{(Newc)}}|>|\omega_{i_{(RN)}}|>|\omega_{i_{(Barc)}}|>|\omega_{i_{(Hawc)}}|$.  
 
The relaxation time is one of the important properties of perturbations and is defined as ($\tau=1/ |\omega_{i}|$). We can mention that the relaxation time of the Newc diminishes faster compared with the Barc and Hawc (see Fig \ref{Fig4} d)), implying that the Newc system recovers the stationary state faster for both perturbations. When we compare the relaxation times for the scalar and  electromagnetic perturbations we can affirm that  $\tau_{(Newc)}<\tau_{(Barc)}<\tau_{(Hawc)}$ for charges $q<0.12$. However, when the magnetic charge approaches its upper bound ($q_{crit}$), the relaxation time grows, which implies that the perturbations in the cases of Barc and Hawc have a longer life.

\section{Reflection and Transmission coefficients}

As the BH absorb matter and fields around them, the accretion rate is created. Accretion has an important role in phenomenology, for example, the study of active galactic nuclei. Another interesting aspect that is the studied is the scattering and absorption cross sections, the latter being related to the shadows of BH (high-frequency limit). As we know, the QNM dominate the last stage of the BH's perturbation process. Moreover, the QNM have an interesting relationship with the scattering and absorption processes of BH.

Due to the aforementioned, in this section, the reflection coefficient and transmission coefficient are discussed for scalar and electromagnetic perturbations of the different classes of black holes.

When an incoming wave towards a black hole is considered, the wave (with a frequency $\omega$) is partially transmitted by the potential (\ref{ec.poseg}) and partially reflected by the same potential (\ref{ec.poseg}). The behavior of the wave can be written as follows;

\begin{eqnarray}
\tilde{\xi}(r_*)&=&T(\omega)e^{-i\omega r_*},\quad r_*\to -\infty\,,\label{xi1} \\
\tilde{\xi}(r_*)&=&e^{-i\omega r_*}+R(\omega)e^{i\omega r_*},\quad r_*\to \infty\,,\label{xi2}
\end{eqnarray}

where $R(\omega)$ represents the reflection coefficient and $T(\omega)$ the transmission coefficient. For the flux to be conserved, the relation between the reflection and transmission is given by;

\begin{equation}\label{ec.cp}
|R(\omega)|^2+|T(\omega)|^2=1\,.
\end{equation}

When the frequency of the incoming wave is minor to the height of the potential barrier ($\omega^{2}\ll V_{max}$), then $T(\omega) \thickapprox 0$ and $R(\omega) \thickapprox 1 $. But, if $\omega^{2}\gg V_{max}$, the reflection coefficient is close to zero, while the transmission coefficient is close to one, then the wave will not be reflected by the barrier of potential.

In the case of $\omega^{2}\approx V_{max}$ the WKB is adequate. The transmission coefficient is understood as the probability that an outgoing wave with a $\omega$  can reach infinity i.e, the absorption probability for an incoming wave to be absorbed. In this case, the wave is absorbed by a BH.

When we consider $\omega^{2}\approx V_{max}$, the reflection coefficient is given by;

\begin{equation}\label{ecu.cr}
	R(\omega)=\left(1+e^{-2\pi i \epsilon}\right)^{-1/2}\,,
\end{equation} 

In third--order in the WKB approximation, $\epsilon$ is determined from the following equation;
\begin{equation}
	\epsilon-i\frac{\omega^2-V_{max}}{\sqrt{-2V^{(2)}_{max}}}=-\Lambda (\epsilon)+\epsilon \Omega (\epsilon)\,,
\end{equation}

where $\Lambda (\epsilon)$ is given by Eq. (\ref{ecu.wkbl}) and $\Omega (\epsilon)$ is given by Eq. (\ref{ecu.wkbo}). Now we can express the transmission coefficient as; 

\begin{equation}
|T(\omega)|^2=1-\left|\left(1+e^{-2\pi i \epsilon}\right)^{-1/2}\right|^2\,.
\end{equation} 

The absorption cross--section can be written as $\widetilde{\sigma}=\sum_{l=0}^{\infty} \sigma_{l}(\omega)$, where the partial absorption cross--section ($\sigma_{l}(\omega)$) of the wave  is given by;

\begin{equation}
	\sigma_{l}(\omega)=\frac{ \pi (2l+1)}{\omega^{2}}|T(\omega)|^2\,.
\end{equation}

\begin{figure}[h]
\begin{center}
\includegraphics [width =0.45 \textwidth ]{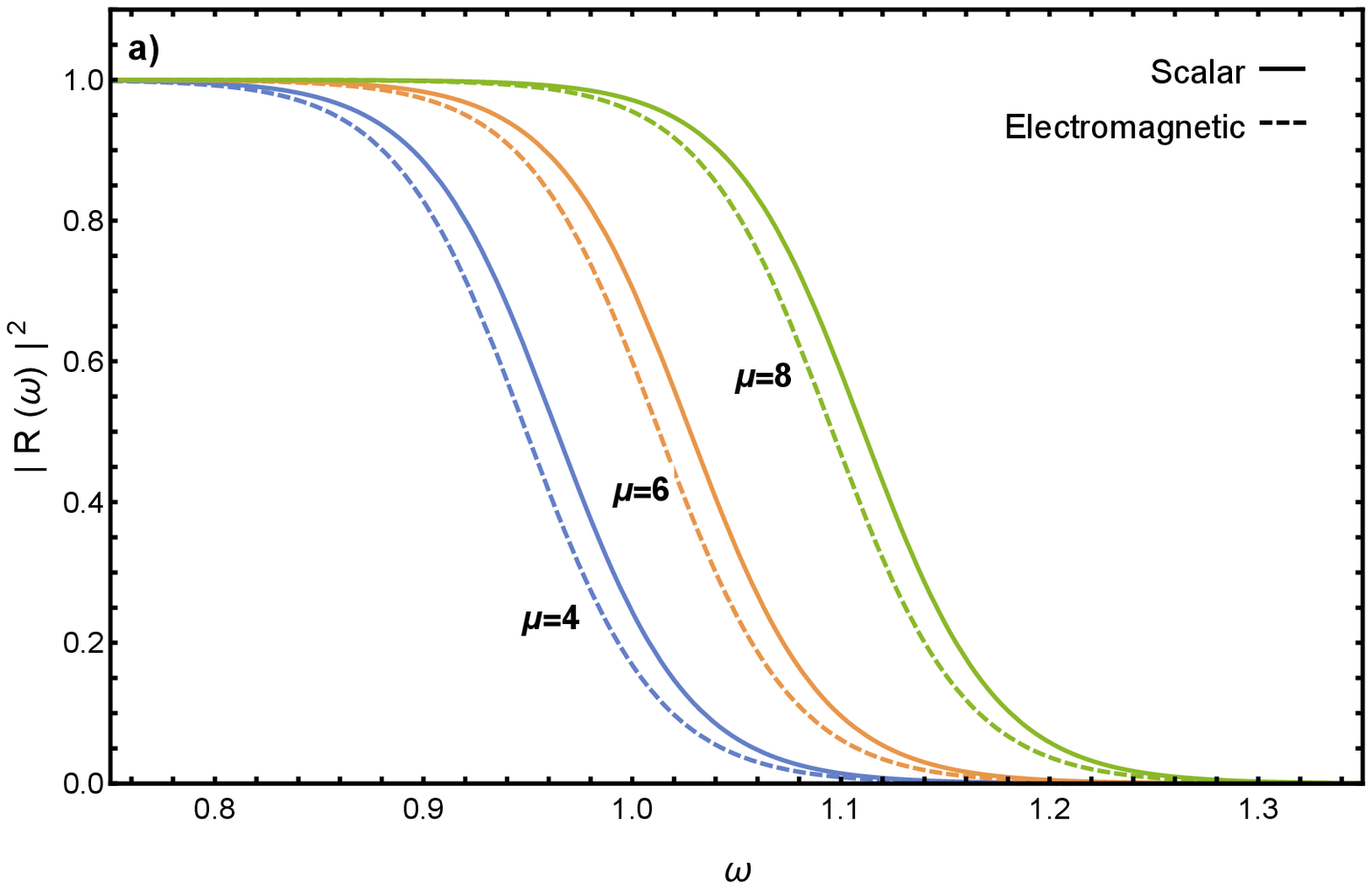}
\includegraphics [width =0.45 \textwidth ]{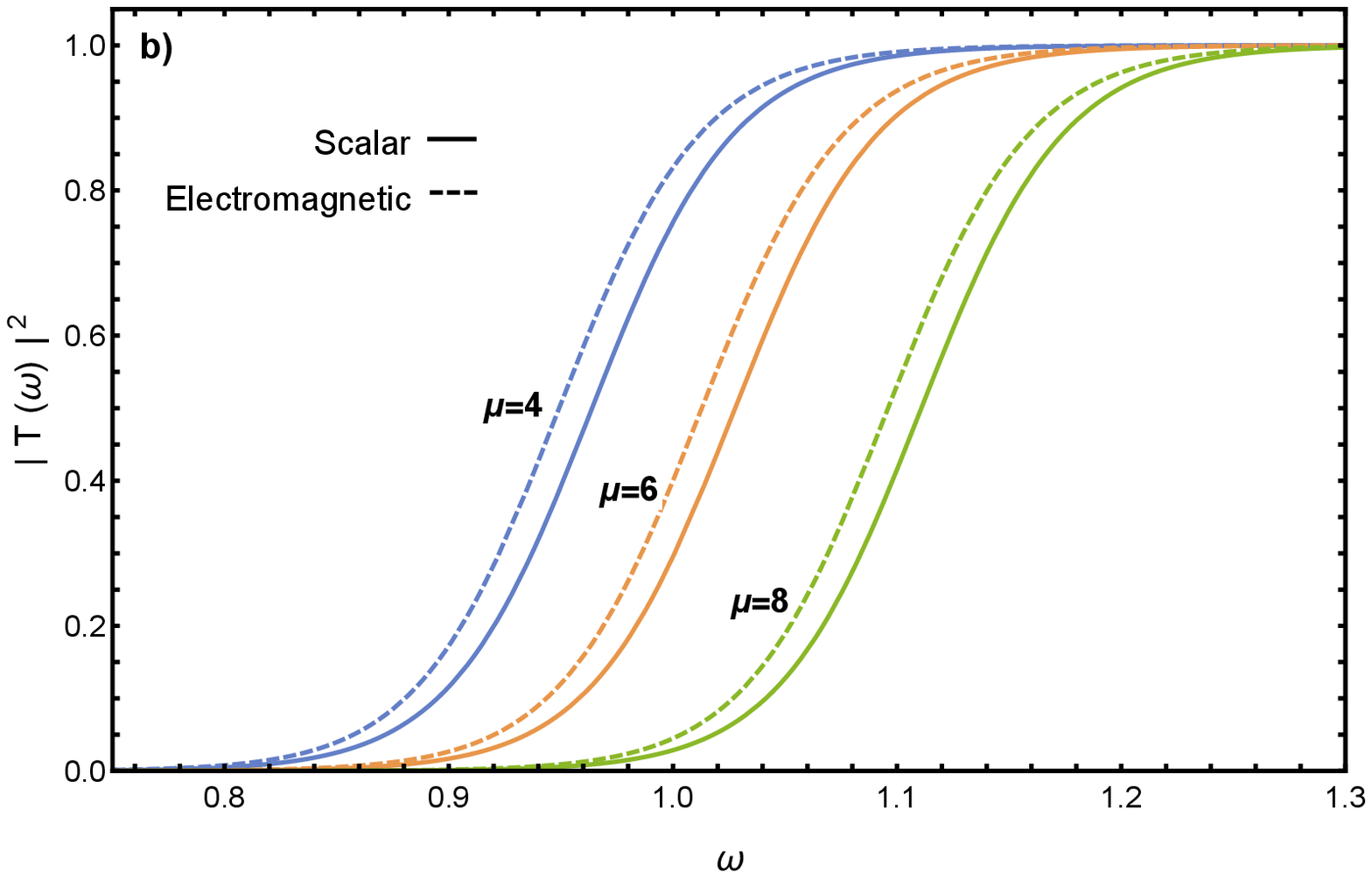}
\end{center}
\caption{The reflection and transmission coefficients for the Newc are shown as a function of $\omega$ with different values of $\mu$, $M=1$, $q=0.07$ and $l=4$.  a) Shows the behavior of $|R(\omega)|^2$.  b) Shows the behavior of $|T(\omega)|^2$.}\label{Fig5}
\end{figure}

\begin{figure}[h]
\begin{center}
\includegraphics [width =0.45 \textwidth ]{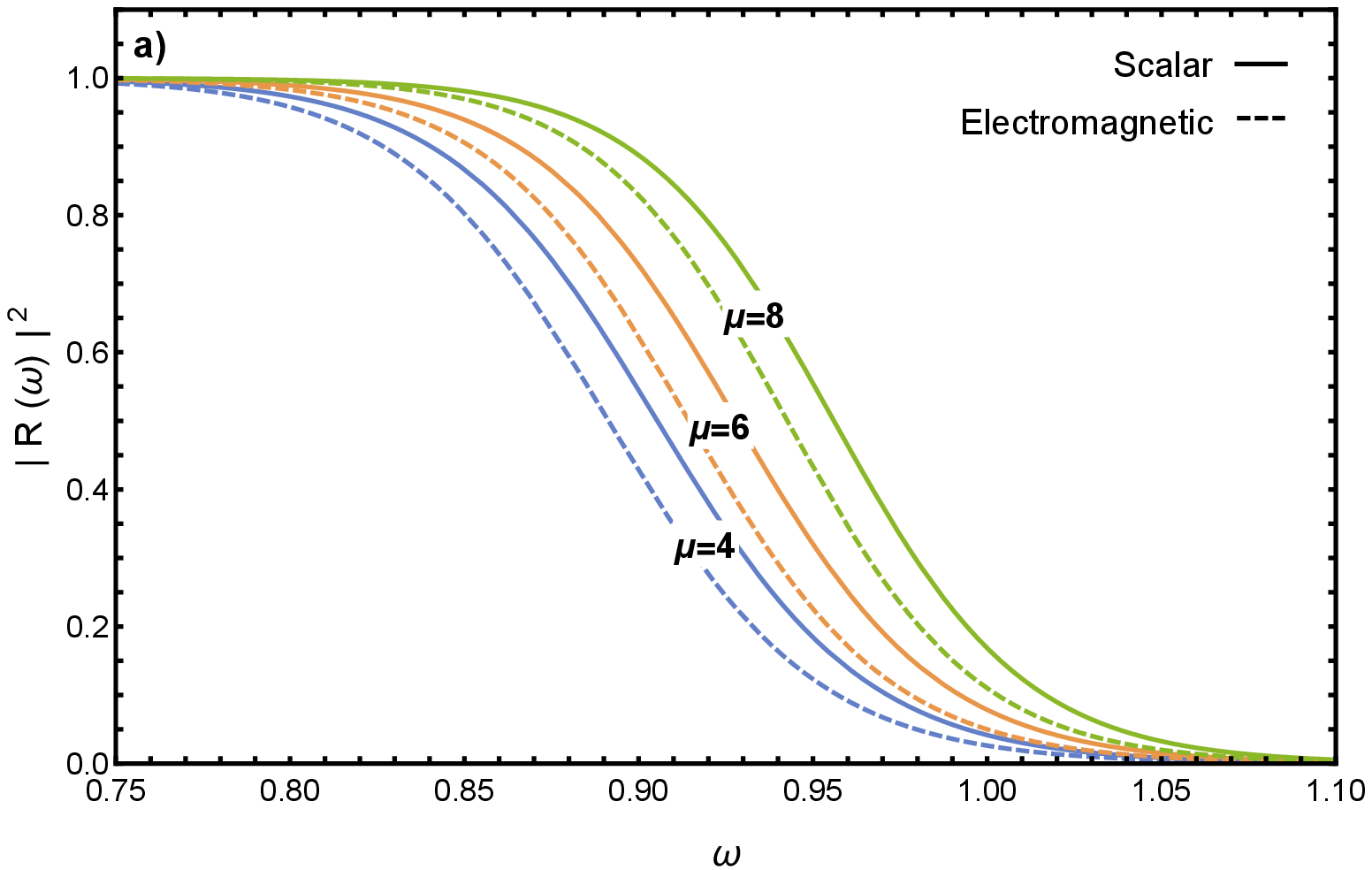}
\includegraphics [width =0.45 \textwidth ]{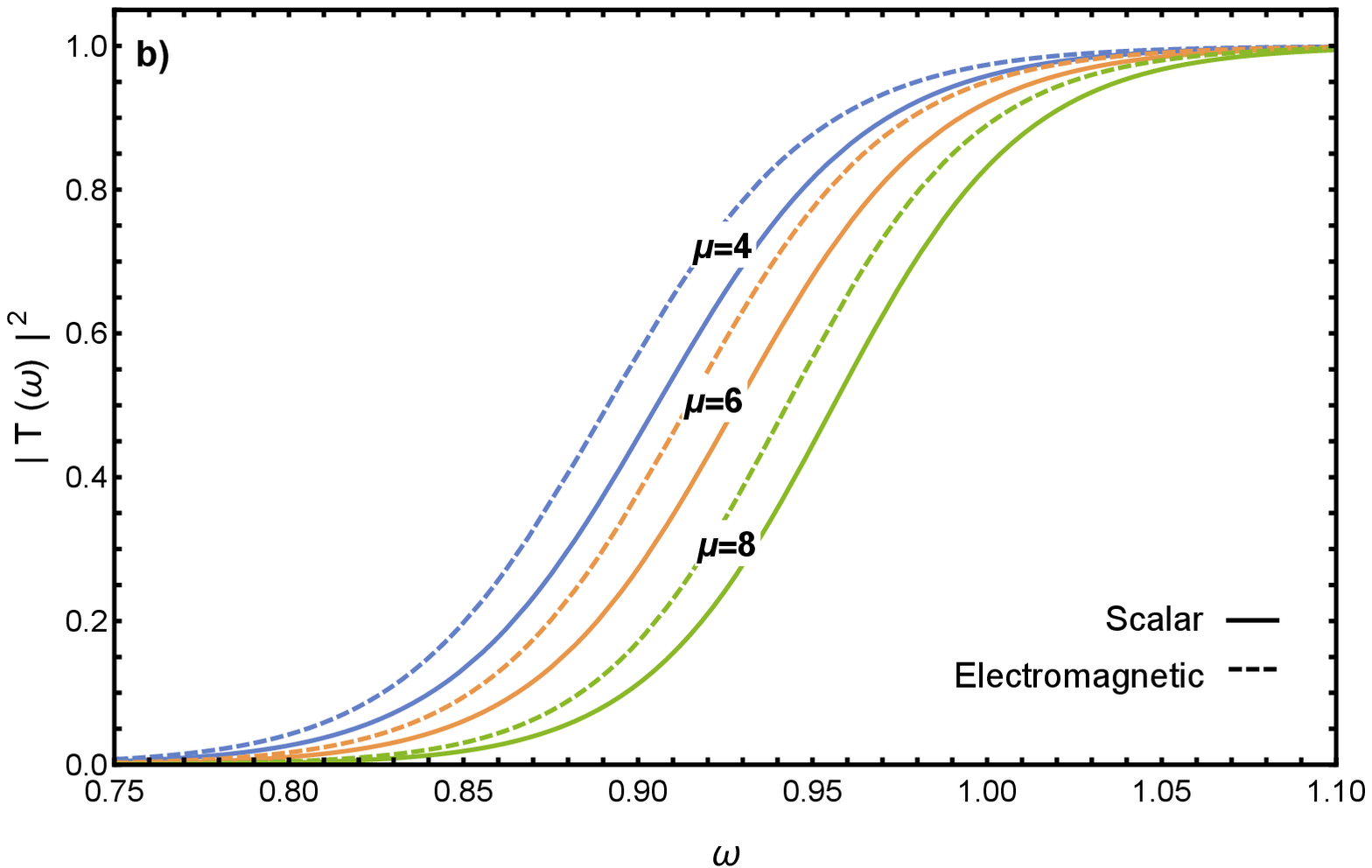}
\end{center}
\caption{The reflection and transmission coefficients for the Barc are shown as a function of $\omega$ with different values of $\mu$, $M=1$, $q=0.41$ and $l=4$.  a) Shows the behavior of $|R(\omega)|^2$.  b) Shows the behavior of $|T(\omega)|^2$.}\label{Fig6}
\end{figure}

\begin{figure}[h]
\begin{center}
\includegraphics [width =0.45 \textwidth ]{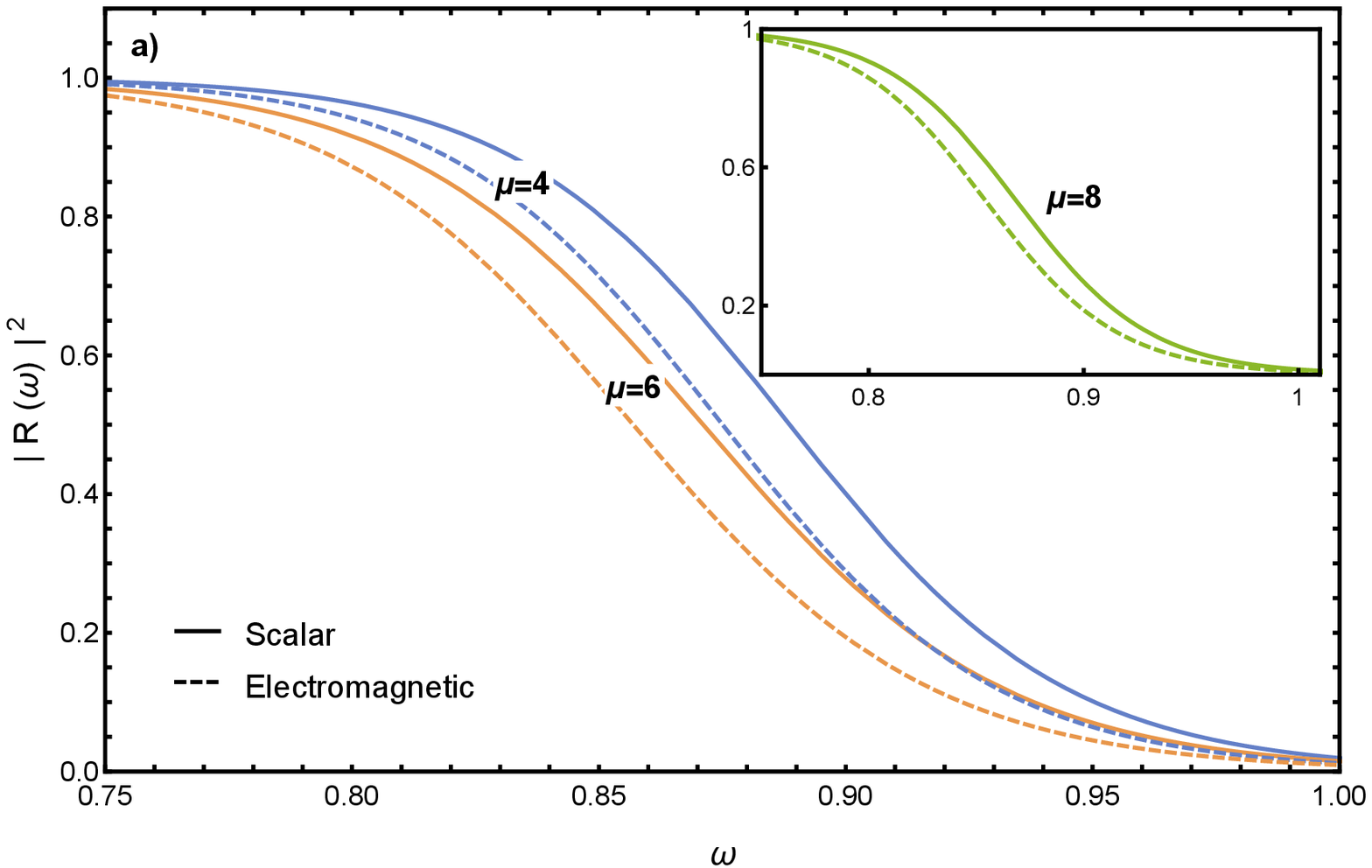}
\includegraphics [width =0.45 \textwidth ]{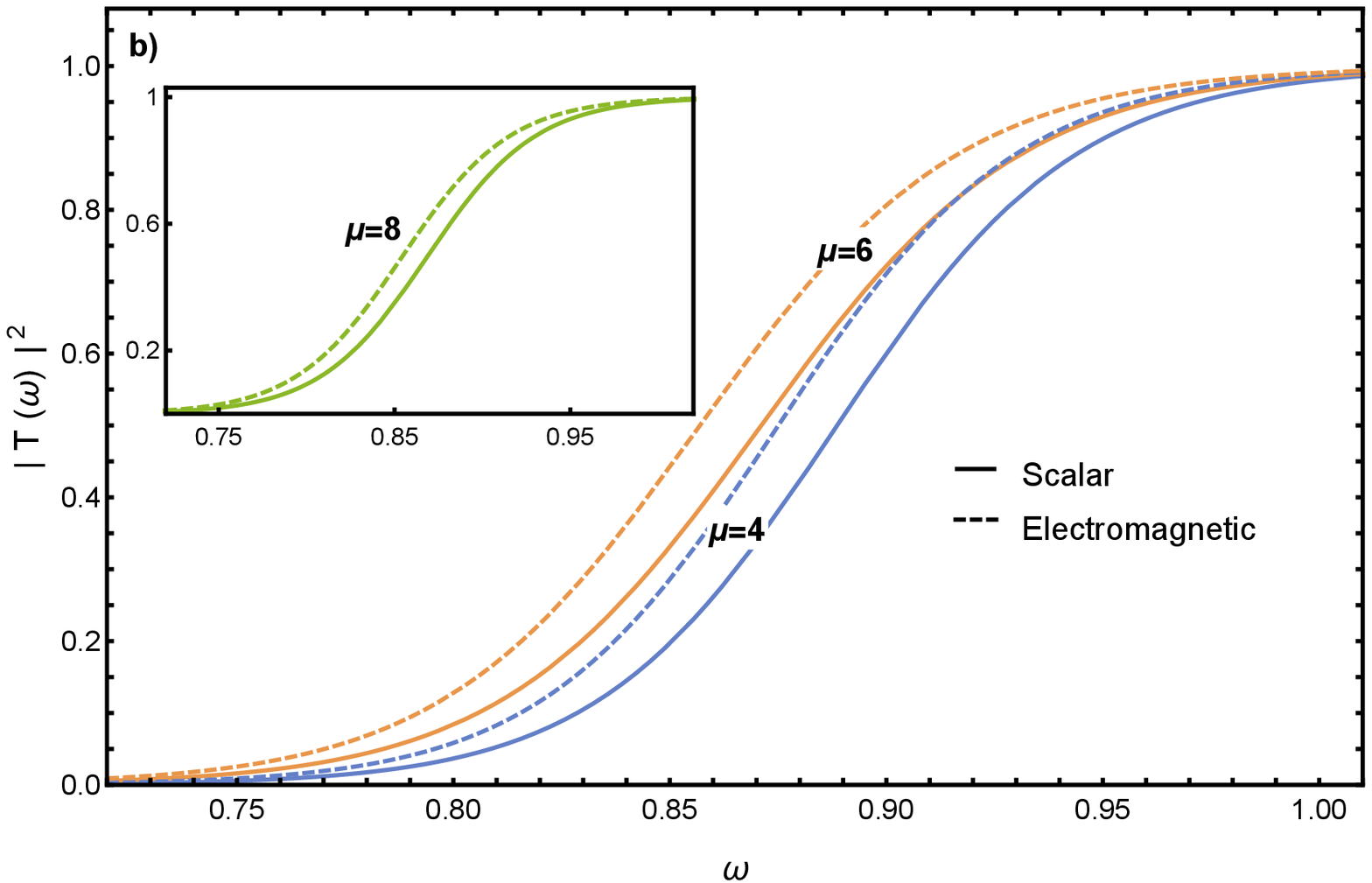}
\end{center}
\caption{The reflection coefficient and transmission coefficient for the Hawc are shown as a function of $\omega$ with different values of $\mu$, $M=1$, $q=1.13$ and $l=4$.  a) Shows the behavior of $|R(\omega)|^2$.  b) Shows the behavior of $|T(\omega)|^2$.}\label{Fig7}
\end{figure}

The reflection and transmission coefficients (with different values of $\mu$) for the Newc, Barc and Newc are shown in Figs. \ref{Fig5}--\ref{Fig7}. Fig. \ref{Fig5} a) shows how the reflection coefficient increases as $\mu$ increases in the case of Newc and the Fig. \ref{Fig5} b) shows the opposite behavior for the  transmission coefficient.
However, the effect is more noticeable in the scalar perturbation than in the case of the electromagnetic perturbation for the reflection coefficient. 

Fig. \ref{Fig6} a) and b) show the reflection coefficient and the transmission coefficient of the Barc for different $\mu$, the differences between the different $|R(\omega)|^2$ or the different $|T(\omega)|^2$, are closer for both perturbations. And as we observe in the Newc the effect of the electromagnetic perturbation is more noticeable.

The reflection and transmission coefficients corresponding to Hawc are shown in Figs. \ref{Fig7} a) and b), as a function of $\omega$ with different values of $\mu$. The reflection coefficient is very close in the cases $\mu=6$ and $\mu=8$. The same happens in the case of the transmission coefficient (it is observed for both perturbations). In the small box, the difference is shown. 

It is clear that the effect of $\mu$ is to diminish the transmission coefficient for the Newc and Barc, but in Hawc it increases. Implying that when the strength of nonlinearity of the electromagnetic field increases the transmission of the wave decreases, for any of the perturbations in Newc and Barc. And, an increase in the strength of nonlinearity of the electromagnetic field increases the transmission for Hawc. In general we can mention that $|R(\omega)|^2 _{Newc}> |R(\omega)|^2 _{Barc}> |R(\omega)|^2 _{RN}>|R(\omega)|^2 _{Hawc}$ and $|T(\omega)|^2 _{Hawc}>|T(\omega)|^2 _{RN}> |T(\omega)|^2 _{Barc}> |T(\omega)|^2 _{Newc}$, when the $\mu$ parameter is fixed and we have the same values of $q$, $M$ and $l$. 

\begin{figure}[h]
\begin{center}
\includegraphics [width =0.45 \textwidth ]{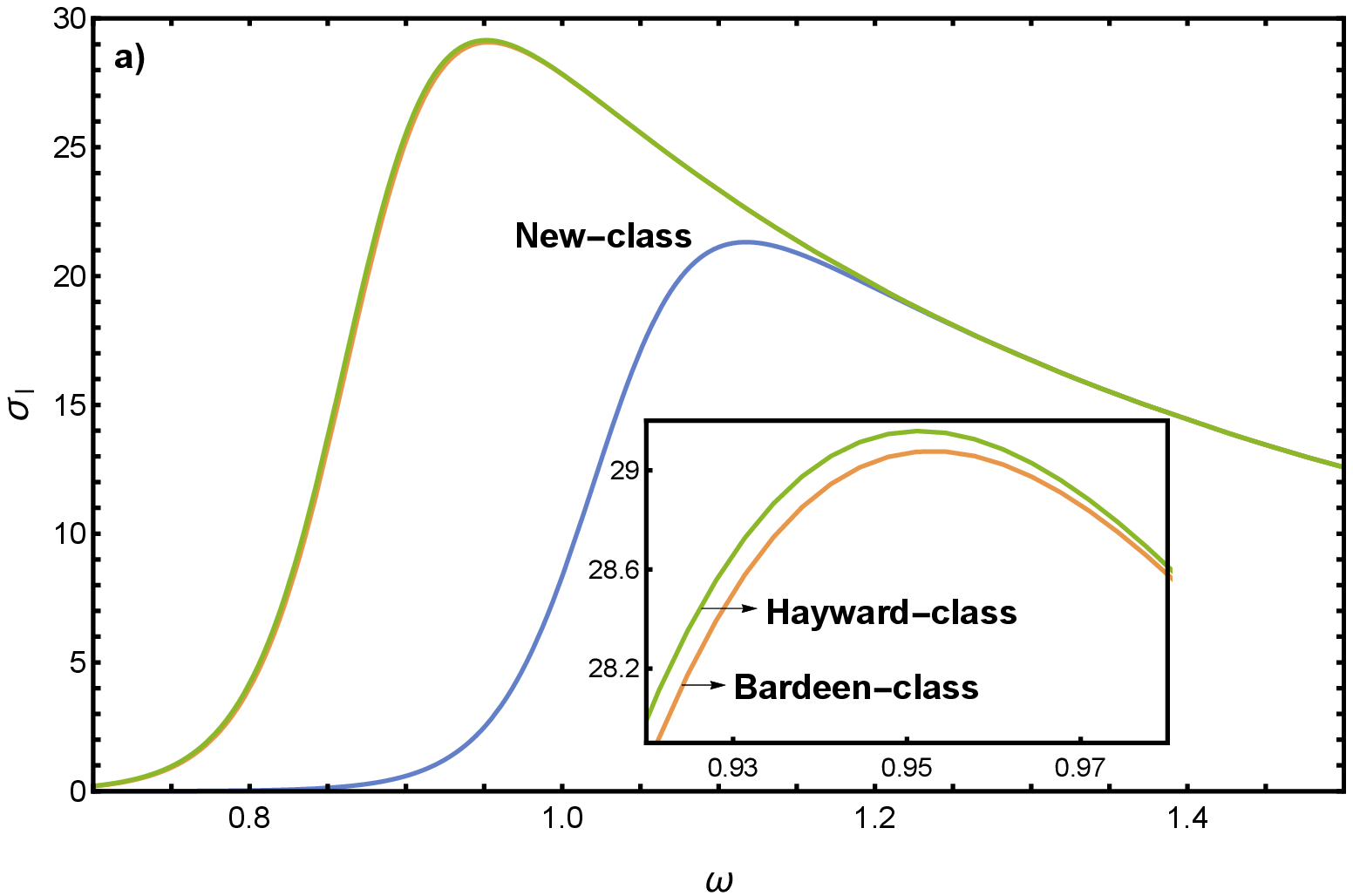}
\includegraphics [width =0.45 \textwidth ]{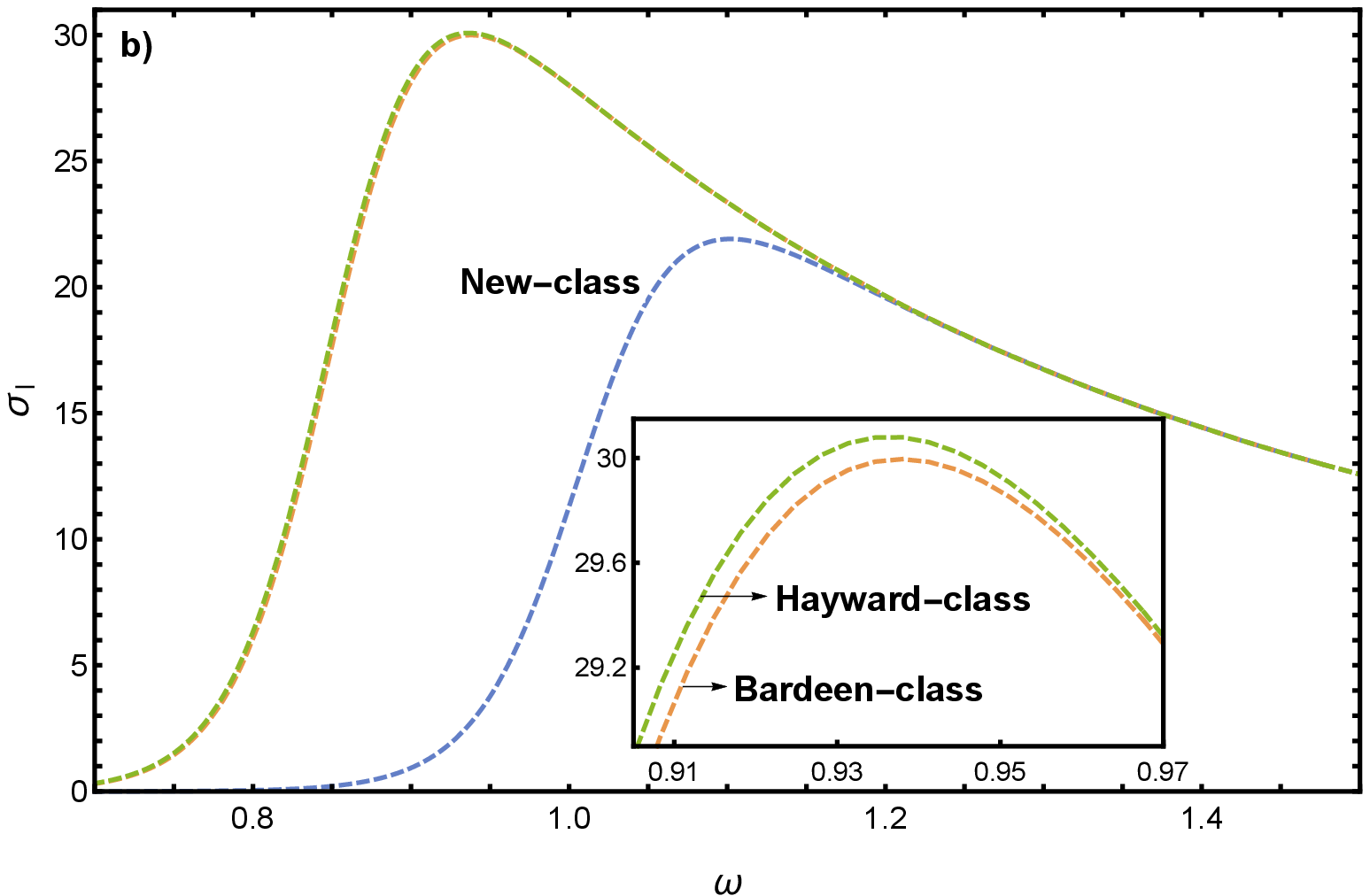}
\end{center}
\caption{The different $\sigma_{l}$ are shown for $l=4$, $\mu=6$, $q=0.07$ and $M=1$. a) The different $\sigma_{l}$ are shown for scattered scalar. b) The different $\sigma_{l}$ are shown for scattered electromagnetic}\label{Fig8}
\end{figure}

Fig. \ref{Fig8} a) and b)  show that the partial absorption cross--sections are higher for Hawc and Barc. Since the heights of the effective potentials in Fig. \ref{Fig2} d) for Hawc and Barc are smaller than the effective potential of Newc. Thus, there is less absorption for the Newc. On the other hand, when we compare the absorption cross-section scalar and electromagnetic, we can conclude that $\sigma_{l(elec)}>\sigma_{l(sc)}$. We can also mention that the absorption cross-section for Hawc is very close to the absorption cross-section of Barc.

\section{Conclusions}
 A Generic--class of regular black hole with a magnetic charge that contains the Barc, Hawc, and  Newc solutions was analized. We studied the event horizons, the extreme case, and presented how magnetic charge ($q$) explicitly depends on $\mu$. The critical mass was also obtained.

Then, by analyzing the effective potential of scalar perturbation and electromagnetic perturbation, it is shown that effective potentials of scalar perturbations are larger compared to the effective potentials of the electromagnetic perturbations in all cases. It is also possible to mention that $V(r)_{Hawc}<V(r)_{Barc}<V(r)_{Newc}$ when we fix $\mu$. When $\mu$ is increased, the maximum of potentials increase for Newc and Barc, but for Hawc it is opposite.

The QNM are studied for the scalar and electromagnetic perturbations in the Generic--class, via the third--order WKB method. We have shown that an increase in $q$ implies a monotonic increase of the $\omega_{r}$ for Newc and Barc. Also, we can see that with an increase in the parameter $\mu$, the real part ($\omega_{r}$) of the QNM frequencies increases. But in the case of Hawc when we increase the parameter $\mu$, $\omega_{r}$ decreases.

We find that in the Generic--class, the imaginary part of the QNM is always larger for the scalar perturbation. However, the roles are swapped for the relaxation time, we also observe that in a certain range $|\omega_{i_{(Newc)}}|>|\omega_{i_{(Barc)}}|>|\omega_{i_{(Hawc)}}|$.  

It is possible to say that as the presence of strength of nonlinearity of the electromagnetic field increases or decreases the magnitudes of oscillation and relaxation times of the different kinds of solutions. Being the Newc, the most stable solution for scalar and electromagnetic perturbation, but the range of the values of $q$ is small.

The reflection and transmission coefficients have been calculated by applying the third--order WKB approach for scalar and electromagnetic perturbation. As a result, the transmission coefficient decreases with an increase of $\mu$ to the Newc and Barc, while for the Hawc it increases with an increase in $\mu$, i.e., the probability of the wave transmission through the potential barrier depends inversely on the maximum of the effective potential. 

In the Hawc, the transmission coefficient increases due to an increase in the value of $\mu$ that weakens the potential barrier. The behavior is oppositely in the cases of Newc and Barc.

\section*{ACKNOWLEDGMENT}

L. A. L\'opez acknowledge the partial financial support of  SNI--CONACYT, M\'exico.

\bibliographystyle{unsrt}

\bibliography{bibliografia}

\end{document}